\documentclass[aps,prb,twocolumn,longbibliography]{revtex4-2}

\usepackage[utf8]{inputenc}
\usepackage{graphicx}
\usepackage{hyperref}
\usepackage{xcolor,comment}
\usepackage{natbib}
\usepackage[T1]{fontenc}

\usepackage{amsmath,amsthm,amssymb}
\usepackage{mathtools}
\usepackage{braket}
\usepackage{bbm,bm}
\usepackage[bb=boondox]{mathalfa}

\newcommand{\ie}{{i.e.},\ }

\begin{document}
\title{Eigenstate thermalization in spin-$\frac{1}{2}$ systems with SU(2) symmetry}

\author{Rohit Patil}
\affiliation{Department of Physics, The Pennsylvania State University, University Park, Pennsylvania 16802, USA}
\author{Marcos Rigol}
\affiliation{Department of Physics, The Pennsylvania State University, University Park, Pennsylvania 16802, USA}

\begin{abstract}
We study the diagonal and off-diagonal matrix elements of observables in the eigenstates of the extended spin-$\frac{1}{2}$ Heisenberg chain, which exhibits the non-Abelian SU(2) symmetry. We explore integrable and nonintegrable regimes, and consider observables that preserve the SU(2) symmetry of the Hamiltonian as well as observables that break it. We study in detail the low-frequency behavior of the off-diagonal matrix elements at and away from integrability. In the nonintegrable regime, we test the non-Abelian eigenstate thermalization hypothesis, paying special attention to the effect of the spin, which is the distinctive conserved quantity introduced by the SU(2) symmetry.      
\end{abstract}

\maketitle

\section{Introduction}
\label{sec:Intro}

Thermalization in generic isolated quantum systems, namely, the fact that after equilibration observables can be described using ensembles of statistical mechanics, has attracted much attention in the last 20 years~\cite{dalessio_quantum_2016, deutsch_18, mori_ikeda_18}. Eigenstate thermalization is the phenomenon behind such a universal behavior of nonintegrable (quantum-chaotic) systems. Its corresponding mathematical ansatz, the eigenstate thermalization hypothesis (ETH)~\cite{deutsch_1991, srednicki_1994, rigol_2008}, states that the matrix elements of observables $\hat O$ in the energy eigenstates of quantum-chaotic Hamiltonians, $\hat H \ket{\psi_\alpha} = E_\alpha \ket{\psi_\alpha}$, have the form: 
\begin{equation}\label{eq:ETH}
 \langle \psi_\alpha|\hat O| \psi_\beta \rangle=O(\bar E)\delta_{\alpha\beta}\,+\,e^{-S_{\text{th}}(\bar E)/2}f_O(\bar E,\omega)R_{\alpha\beta}\,,
 \end{equation}
where $\bar E=(E_\alpha+E_\beta)/2$ is the average energy, $\omega=E_\alpha-E_\beta$ is the energy difference, $O(\bar E)$ and $f_O(\bar E,\omega)$ are smooth functions, $S_{\text{th}}(\bar E)$ is the thermodynamic entropy at the energy $\bar E$, and $R_{\alpha\beta}$ are Gaussian random numbers with zero mean and unit variance (variance 2) for $\alpha\neq \beta$ ($\alpha=\beta$) in systems with time-reversal symmetry. The function $O(\bar E)$ is the thermal expectation value of the observable $\hat O$ at the energy $\bar E$ predicted by the microcanonical ensemble, while the function $f_O(\bar E,\omega)$ encodes quantum fluctuations and all the dynamical information from the energy eigenstates~\cite{dalessio_quantum_2016}. The ETH~\eqref{eq:ETH} has been extensively tested in numerical studies~\cite{rigol_2008, rigol_09a,*rigol_09b, kim_testing_ETH_2014, beugeling_2014,*beugeling_offdiag_2015, mondaini_2016, mondaini_2017, leblond_2019, jansen_2019, leblond_2020, schonle_autocorrelation_2021, wang_2024} (see Refs.~\cite{dalessio_quantum_2016, deutsch_18, mori_ikeda_18} for reviews).

As written in Eq.~\eqref{eq:ETH}, the ETH applies to quantum systems in the absence of symmetries. Once symmetries are present, one needs to resolve them, and Eq.~\eqref{eq:ETH} applies to irreducible sectors of the spectrum of the Hamiltonian. The most common symmetries in lattice models of interest in condensed matter physics are discrete position-space symmetries such as lattice translations and reflections (parity), and continuous Abelian symmetries such as U(1) symmetry which leads to magnetization conservation in spin systems (particle-number conservation in itinerant models). If those symmetries are not taken into account, indicators of quantum chaos such as the level-spacing statistics cannot be used to distinguish chaotic from non-chaotic systems~\cite{dalessio_quantum_2016, santos_rigol_10a,*santos_rigol_10b}. The type of symmetry has important consequences on the ETH across symmetry sectors, e.g., for discrete lattice symmetries (which are nonlocal and do not have an associated extensive conserved quantity) the diagonal ETH has been found to be the same across symmetry sectors~\cite{rigol_09a,*rigol_09b, mondaini_mallayya_18}, while this is generally not the case for the U(1) symmetry. In the presence of U(1) symmetry, the diagonal ETH acquires a dependence on the magnetization (which is the associated extensive conserved quantity) of the sector considered.

Novel to the traditional way of thinking about the ETH is the presence of non-Abelian symmetries such as the SU(2) symmetry, which is also common in spin models of interest in condensed matter physics. When $\hat H$ exhibits a non-Abelian symmetry, it does not share simultaneous eigenstates with all the noncommuting conserved quantities, i.e., not all conserved quantities can be simultaneously resolved to use the ETH as per Eq.~\eqref{eq:ETH}. The study of systems with noncommuting conservation laws has garnered significant attention in the context of quantum information thermodynamics~\cite{halpern_2016, guryanova_2016, popescu_2020, halpern_2020, manzano_2022, kranzl_2023, majidy_review_2023} and, recently, the effect of noncommuting conservation laws on the entanglement entropy of energy eigenstates~\cite{patil_2023, majidy_entanglement_2023, bianchi_2024} and on the ETH~\cite{murthy_nonabelian_2023, noh_2023, lasek_24} has begun to be studied in detail. On the experimental side, thermalization close to a ``non-Abelian thermal state''~\cite{halpern_2016} was observed in a trapped ion quantum simulator~\cite{kranzl_2023}. 

The goal of this paper is to deepen our understanding of the effect of non-Abelian symmetries on the matrix elements of observables in the energy eigenstates. Specifically, we consider the extended spin-$\frac{1}{2}$ Heisenberg chain, which is SU(2) symmetric, and study observables that preserve and break the SU(2) symmetry of the Hamiltonian. Our main interest is eigenstate thermalization (a phenomenon that occurs in the nonintegrable regime) in the presence of SU(2) symmetry, and we contrast it against results for the integrable spin-$\frac{1}{2}$ Heisenberg chain. We pay special attention to the effect of the spin on the expectation values of observables in the energy eigenstates, and study in detail the low-frequency behavior of the off-diagonal matrix elements of the observables.

The presentation is organized as follows. In Sec.~\ref{sec:Model}, we introduce the extended spin-$\frac{1}{2}$ Heisenberg chain and the observables studied, and connect the matrix elements of the latter in the energy eigenstates to the non-Abelian ETH proposed in Ref.~\cite{murthy_nonabelian_2023}. We report our results for the diagonal ETH, which probe both the energy and spin dependence of the diagonal matrix elements of observables, in Sec.~\ref{sec: Diag}.  Section~\ref{sec:Off-diag} is devoted to the study of the off-diagonal ETH. Specifically, in Sec.~\ref{sec:Off-diag-equalspins} (Sec.~\ref{sec:Off-diag-unequalspins}), we report our results for the off-diagonal ETH for observables that connect energy eigenstates with the same spin (different spins). We study in detail the properties and scalings of the off-diagonal matrix elements, and the low- and high-frequency $\omega$ dependence of the spectral function. We conclude with a summary and discussion of the results in Sec.~\ref{sec: Summary}.

\section{Model}\label{sec:Model}

We study the extended spin-$\frac{1}{2}$ Heisenberg model with nearest and next-nearest neighbor interactions in chains with $L$ sites and periodic boundary conditions    
\begin{equation}\label{eq:HamiltonianSpin1/2}
    \hat H=-\sum_{i=1}^L \hat{\vec{S}}_i \cdot \hat{\vec{S}}_{i+1} - \lambda \sum_{i=1}^L  \hat{\vec{S}}_i \cdot \hat{\vec{S}}_{i+2}\,,  
\end{equation}
where $\hat{\vec{S}}_i=(\hat{S}^x_i,\hat{S}^y_i,\hat{S}^z_i)$ is the spin-$\frac12$ operator at site $i$. This model is integrable for $\lambda=0$ and nonintegrable otherwise. We set $\lambda=3$ to study the latter regime~\cite{patil_2023}.

The Hamiltonian~\eqref{eq:HamiltonianSpin1/2} is SU(2) symmetric, \ie the total spin $S$ and the total magnetization $M$, corresponding to the eigenvalues $S(S+1)$ and $M$ of $\hat{\vec{S}}^2=(\sum_i \hat{\vec{S}}_i)^2$ and $\hat M=\sum_i \hat{S}^z_i$, respectively, are conserved (we set $\hbar=1$). We take $L$ to be even and focus on the zero total magnetization sector ($M=0$). The Hamiltonian~\eqref{eq:HamiltonianSpin1/2} is translationally invariant, so we block-diagonalize it in quasi-momentum space. For $L$ even, the quasimomentum $k=\frac{2\pi n}{L}$ where $n\in \{-L/2+1,-L/2+2,\ldots,L/2\}$. We exclude the $k=0$ and $\pi$ quasi-momentum sectors from our calculations because they exhibit parity symmetry, which results in stronger finite-size effects compared to all other $k\neq0,\pi$ sectors~\cite{leblond_2020}. Furthermore, for $M=0$, there is spin-inversion symmetry along the $z$-axis ($Z_2$), which partitions each $k$ sector into two $Z_2$ subsectors. We therefore use exact diagonalization in the simultaneous $\{M,k,Z_2\}$ basis with $M=0$, $k\neq0,\pi$, and $Z_2=\pm1$ to obtain the energy eigenstates for the Hamiltonian. Lastly, we resolve the total spin $S$ by direct calculation of the eigenvalue of $\hat{\vec{S}}^2$ in each energy eigenstate. 

In systems with SU(2) symmetry, it is convenient to decompose observables in terms of spherical tensor operators $\hat T^{(r)}_q$, where $r$ is the rank of the tensor with components $q=-r,\,-r+1,\,\ldots,\,r$, which transform irreducibly under rotations~\cite{Shankar1994, Sakurai2017}. The spherical tensor operators $\hat T^{(r)}_q$ form a basis for all operators, $\hat O=\sum_{r,q} C_{r,q}\hat T^{(r)}_q$. For instance, the $z$-component nearest-neighbor interaction takes the form $\hat S^z_1\hat S^z_2=-\sqrt{\frac{1}{3}}\hat T^{(0)}_0+\sqrt{\frac{2}{3}}\hat T^{(2)}_0$, where $\hat T^{(0)}_0=-\sqrt{\frac{1}{3}}\hat{\vec{S}}_1 \cdot \hat{\vec{S}}_2$ and $\hat T^{(2)}_0=\sqrt{\frac{1}{6}}(3\hat S^z_1\hat S^z_2 -\hat{\vec{S}}_1 \cdot \hat{\vec{S}}_2)$. The operators $\hat T^{(r)}_q$ are useful as their matrix elements in the simultaneous eigenstates \{$|E_\alpha S_\alpha M_\alpha\rangle$\} of $\hat{H}$, $\hat{\vec{S}}^2$, and $\hat M$ can be written (using the Wigner-Eckart theorem) as
\begin{align}
    \langle E_\alpha S_\alpha M_\alpha|\hat T^{(r)}_q|E_\beta S_\beta M_\beta\rangle=&\langle S_\alpha M_\alpha|S_\beta M_\beta; r q\rangle \,\label{eq:Wigner-EckartTheorem} \\ 
    &\times \langle E_\alpha S_\alpha||T^{(r)}||E_\beta S_\beta\rangle\,,\nonumber 
\end{align}
where $\langle S_\alpha M_\alpha|S_\beta M_\beta;r q\rangle$ is the Clebsch-Gordan coefficient $\langle jm|j_1m_1;j_2m_2\rangle$ corresponding to the addition of angular momenta $j_1$ and $j_2$ to give a total angular momentum $j$. This coefficient satisfies the angular momentum addition: $M_\alpha=M_\beta+q$ and $|S_\beta-r|\leqslant S_\alpha \leqslant S_\beta+r$. All the information that is independent of $M_\alpha,\, M_\beta$, and $q$ in the matrix elements of $\hat T^{(r)}_q$ is encoded in the reduced matrix element $\langle E_\alpha S_\alpha||T^{(r)}||E_\beta S_\beta\rangle$. Computing the matrix elements of $\hat T^{(r)}_q$ for a given choice of $M_\alpha,\,M_\beta$, and $q$ (with a non-zero Clebsch-Gordan coefficient) allows one to obtain $\langle E_\alpha S_\alpha||T^{(r)}||E_\beta S_\beta\rangle$, which in turn can be used to determine the matrix elements of $\hat T^{(r)}_q$  for all the other choices of $M_\alpha,\,M_\beta$, and $q$. 

The non-Abelian ETH, as postulated in Ref.~\cite{murthy_nonabelian_2023}, is an ansatz for the reduced matrix elements $\langle E_\alpha S_\alpha||T^{(r)}||E_\beta S_\beta\rangle$:
\begin{align}
    \langle E_\alpha S_\alpha||T^{(r)}||E_\beta S_\beta\rangle&=T^{(r)}(\bar E,\bar S)\delta_{\alpha\beta}\, \label{eq:non-AbelianETH}  \\
     &+\, e^{-S_{\text{th}}(\bar E,\bar S)/2}f^{(r)}_{T}(\bar E,\omega;\bar S,\nu)R_{\alpha\beta}\,.\nonumber
\end{align}
In Eq.~\eqref{eq:non-AbelianETH}, the smooth function $T^{(r)}(\bar E,\bar S)$ and thermodynamic entropy $S_{\text{th}}(\bar E,\bar S)$ depend on the average energy $\bar E=(E_\alpha+E_\beta)/2$ and the average spin $\bar S=(S_\alpha+S_\beta)/2$, while the smooth function $f^{(r)}_{T}(\bar E,\omega;\bar S,\nu)$ additionally depends on the energy and spin differences $\omega=E_\alpha-E_\beta$ and $\nu=S_\alpha-S_\beta$, respectively.

The form of the ansatz in Eq.~\eqref{eq:non-AbelianETH} is the one expected to apply to observables in systems in which there is one extensive conserved quantity in addition to the energy, e.g., in the presence of U(1) symmetry: 
\begin{align}
    \langle E_\alpha M_\alpha|\hat O|E_\beta M_\beta\rangle&=O(\bar E,\bar M)\delta_{\alpha\beta}\, \label{eq:gen-AbelianETH} \\
     &+\, e^{-S_{\text{th}}(\bar E,\bar M)/2}f_O(\bar E,\omega;\bar M,\nu)R_{\alpha\beta}\,,\nonumber
\end{align}
where $\bar M=(M_\alpha+M_\beta)/2$ is the average magnetization and $\nu=M_\alpha-M_\beta$ is the corresponding magnetization difference.

In this paper we study the matrix elements of two observables that are spherical tensor operators: a rotationally invariant observable $\hat A$ (of the form $\hat T^{(0)}_0$) and a non-rotationally invariant observable $\hat B$ (of the form $\hat T^{(2)}_0$):
\begin{eqnarray}\label{eq:observableA}
    \hat A&=&\frac{-1}{\sqrt{3}L}\sum_{i=1}^L \hat{\vec{S}}_i \cdot \hat{\vec{S}}_{i+1}\,, \\ \label{eq:observableB}
    \hat B&=&\frac{1}{\sqrt{6}L}\sum_{i=1}^L (3\hat S_i^z \hat S_{i+1}^z-\hat{\vec{S}}_i \cdot \hat{\vec{S}}_{i+1})\,.
\end{eqnarray}
The observable $\hat A$ is proportional to the Hamiltonian~\eqref{eq:HamiltonianSpin1/2} in the integrable limit ($\lambda=0$). It is a scalar SU(2) symmetric spherical tensor of the form $\hat T^{(0)}_0$, which can only connect energy eigenstates $\langle E_\alpha S_\alpha M_\alpha|\hat A|E_\beta S_\beta M_\beta\rangle$ with the same spin $S_\alpha=S_\beta$ and magnetization $M_\alpha=M_\beta$. On the other hand, the observable $\hat B$ is a $\hat T^{(2)}_0$ operator corresponding to the quadrupole moment, which breaks the SU(2) symmetry of the Hamiltonian. Therefore, it can connect states $\langle E_\alpha S_\alpha M_\alpha|\hat B|E_\beta S_\beta M_\beta\rangle$ with the same magnetization $M_\alpha=M_\beta$ and with spins $|S_\beta-2|\leqslant S_\alpha \leqslant S_\beta+2$.      

We compute the diagonal and off-diagonal matrix elements of the two observables $\hat O=\hat A,\hat B$,   
\begin{equation}\label{eq:Matrixelement}
    O_{\alpha\beta}\equiv \langle E_\alpha S_\alpha M_\alpha|\hat O|E_\beta S_\beta M_\beta\rangle\,,    
\end{equation}
with respect to the energy eigenstates within a simultaneous symmetry sector labeled by $\{S,M,k,Z_2\}$. Focusing on the $M=0$ sector, the results reported for a fixed spin $S$ (and hence also fixed $Z_2$) are computed as a weighted (by the Hilbert space dimension) average over all the $k\neq0,\pi$ total quasimomentum sectors.

Since we consider energy eigenstates that have a fixed zero total magnetization $M_\alpha=M_\beta=0$ and only the total spin is changed, in analogy with Eq.~\eqref{eq:gen-AbelianETH} for a single conserved quantity we can write
\begin{equation}
    O_{\alpha\beta}=O(\bar E,\bar S)\delta_{\alpha\beta}+ e^{-S_{\text{th}}(\bar E,\bar S)/2}f_O(\bar E,\omega;\bar S,\nu)R_{\alpha\beta}\,.\label{eq:Fullmatrixelementnon-AbelianETH}
\end{equation}
Furthermore, since our observables are spherical tensor operators, the matrix elements $O_{\alpha\beta}\equiv (T^{(r)}_q)_{\alpha\beta}$ are related to the reduced matrix elements $\langle E_\alpha S_\alpha||T^{(r)}||E_\beta S_\beta\rangle$ appearing in the non-Abelian ETH in Eq.~\eqref{eq:non-AbelianETH} via a simple rescaling using the Clebsch-Gordan coefficient $\langle S_\alpha M_\alpha|S_\beta M_\beta; r q\rangle$ according to the Wigner-Eckart theorem~\eqref{eq:Wigner-EckartTheorem}. Therefore, if Eq.~\eqref{eq:Fullmatrixelementnon-AbelianETH} describes our results, then Eq.~\eqref{eq:non-AbelianETH} also does.

\section{Diagonal matrix elements}
\label{sec: Diag}

We first study the diagonal matrix elements of the two observables $\hat A$ and $\hat B$ in energy eigenstates with vanishing and nonvanishing (as $L$ increases) spin densities ($s= S/L$). For the vanishing spin density case, we take $S=0$ ($S=1$) for observable $\hat A$ ($\hat B$~\footnote{Since $\hat B$ is a $\hat T^{(2)}_0$ operator, its diagonal matrix elements vanish between states with $S=0$. Therefore, we fix $S=1$ so that the diagonal matrix elements are nonvanishing yet, in the limit $L = \infty$, the spin density $s = 0$ as for $S=0$ considered for observable $\hat A$.}). For the nonvanishing spin density case, we take $S=L/4$ ($s=1/4$) for both observables.

\begin{figure}[!t]
    \includegraphics[width=0.98\columnwidth]{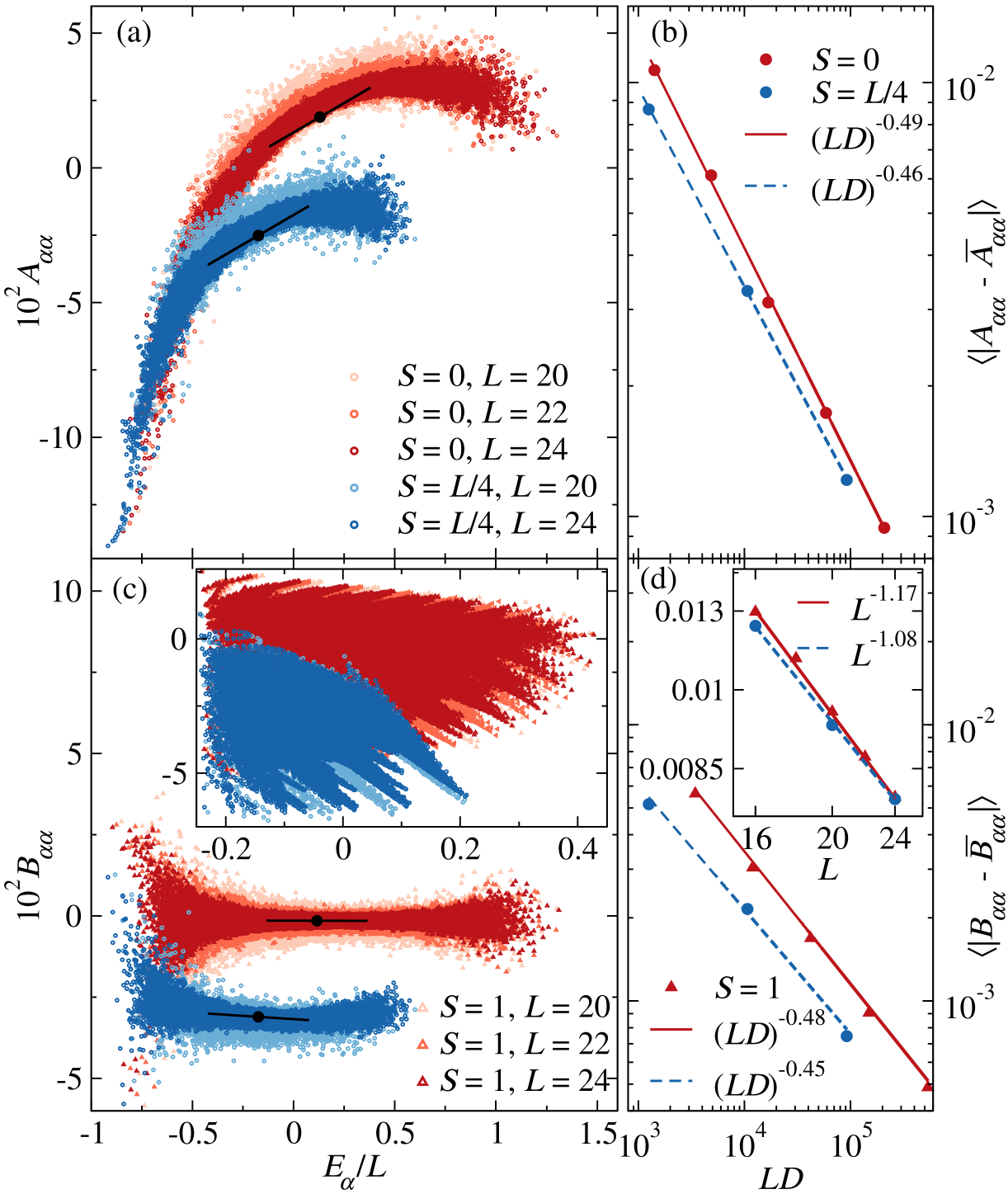}
    \vspace{-0.2cm}
    \caption{Left column: Diagonal matrix elements of observables $\hat A$ and $\hat B$ vs the energy density for different values of $s$ and different system sizes. (a) Matrix elements of $\hat A$ for $S=0$ and $S=L/4$, and (c) matrix elements of $\hat B$ for $S=1$ and $S=L/4$, in the nonintegrable regime with $\lambda=3$. The inset in (c) shows the corresponding matrix elements of $\hat B$ at integrability ($\lambda=0$). Data points with darker shades depict results for larger system sizes. Right column: Corresponding fluctuations of the diagonal elements $O_{\alpha\alpha}$ about the mean $\overline{O_{\alpha\alpha}}$ (computed as a running average over 50 states centered at $E_\alpha$) averaged $\langle\ldots\rangle$ over the central $50\%$ of the spectrum for $\hat O=\hat A$ (b) and $\hat O= \hat B$ (d) in the nonintegrable regime. The inset in (d) shows the results for $\hat B$ at integrability. The results for $S=0,\,1$ ($S=L/4$) were obtained for chains with $L=16,\, 18,\, \ldots,\,24$ ($L=16,\, 20,\, 24$). For $L=24$ in (a) and (c), the infinite-temperature mean values [see Eqs.~\eqref{eq:<A>} and~\eqref{eq:<B>}] are shown as black dots and the linear in energy density terms with the coefficients from Eqs.~\eqref{eq:linear-coeff-A} and \eqref{eq:linear-coeff-B} are shown as black lines centered at the black dots.}
    \label{fig:DiagonalvsEnergydensity}
\end{figure}

Figures~\ref{fig:DiagonalvsEnergydensity}(a) and~\ref{fig:DiagonalvsEnergydensity}(c) show the diagonal matrix elements of $\hat A$ and $\hat B$, respectively, vs the energy density $E_\alpha/L$ for the Hamiltonian~\eqref{eq:HamiltonianSpin1/2} in the nonintegrable regime ($\lambda =3$). For both the vanishing (red) and nonvanishing (blue) spin densities, the support of the matrix elements shrinks rapidly with increasing system size indicating that the system exhibits eigenstate thermalization. In the integrable regime ($\lambda=0$), on the other hand, the inset in Fig.~\ref{fig:DiagonalvsEnergydensity}(c) shows that the support of $B_{\alpha\alpha}$ does not decrease with increasing system size indicating that the system does not exhibit eigenstate thermalization. 

In the nonintegrable regime, the smooth function $O(E_\alpha,S_\alpha)$ [see Eq.~\eqref{eq:Fullmatrixelementnon-AbelianETH}] within a sector with total spin $S$ and magnetization $M=0$ can be written, about the mean energy at infinite temperature $E_0=\braket{\hat H}_{S,M=0}=\mathrm{Tr}(\hat H)/\mathrm{Tr}(1)$, as
\begin{equation}\label{eq:Taylor-expansion}
    O(E_\alpha,S)\approx O(E_0,S)+\frac{\partial O(E_\alpha,S)}{\partial (E_\alpha/L)}|_{E_\alpha=E_0}\big(\frac{E_\alpha-E_0}{L}\big)+\ldots\,,
\end{equation}
where $O(E_0,S)=\braket{\hat O}_{S,M=0}=\mathrm{Tr}(\hat O)/\mathrm{Tr}(1)$ is the infinite temperature expectation value of $\hat O$. The coefficient of the linear in energy density term is given by~\cite{capizzi_poletti_24}
\begin{equation}\label{eq:linear-coeff}
    \frac{\partial O(E_\alpha,S)}{\partial (E_\alpha/L)}|_{E_\alpha=E_0}=L\frac{\braket{\hat O\hat H}_c}{\braket{\hat H \hat H}_c}\,,
\end{equation}
where $\braket{\hat X \hat H}_c=\braket{\hat X \hat H}_{S,M=0}-\braket{\hat X}_{S,M=0}\braket{\hat H}_{S,M=0}$ is the joint connected cumulant between $\hat X$ and $\hat H$. The diagonal functions for the observables $\hat A$ and $\hat B$ for energies $E_\alpha$ close to $E_0$, within the total spin $S$ and total magnetization $M=0$ sector, are given by Eq.~\eqref{eq:Taylor-expansion} with the corresponding infinite temperature mean values
\begin{align}\label{eq:<A>}
    A(E_0,S)&=\frac{1}{\sqrt{3}}\frac{\frac{3}{4}L-S(S+1)}{L(L-1)}\,, \\
    B(E_0,S)&=\frac{-1}{\sqrt{6}}\frac{S(S+1)}{L(L-1)}\,,\label{eq:<B>}
\end{align}
and the linear coefficients
\begin{align}\label{eq:linear-coeff-A}
    \frac{\partial A(E_\alpha,S)}{\partial (E_\alpha/L)}|_{E_\alpha=E_0}&=\frac{L-9}{2\sqrt{3}(5L-21)}\,, \\
    \frac{\partial B(E_\alpha,S)}{\partial (E_\alpha/L)}|_{E_\alpha=E_0}&=\frac{-\sqrt{\frac{2}{3}}S(S+1)(L-9)}{[3(L-2)^2-4S(S+1)](5L-21)}\,,\label{eq:linear-coeff-B}
\end{align}
see Appendix~\ref{app:Infinite-temperature-expectation-values}.

Figures~\ref{fig:DiagonalvsEnergydensity}(a) and~\ref{fig:DiagonalvsEnergydensity}(c) show the leading order energy dependence of $A(E_\alpha,S)$ and $B(E_\alpha,S)$ for the largest chain ($L=24$), with the infinite-temperature mean values from Eqs.~\eqref{eq:<A>} and~\eqref{eq:<B>} indicated by black dots and the linear in energy density terms with the coefficients from Eqs.~\eqref{eq:linear-coeff-A} and \eqref{eq:linear-coeff-B} shown as black lines. For both observables in the thermodynamic limit, the mean values vanish for vanishing spin densities while they approach an (spin-density-dependent) $O(1)$ number for nonvanishing spin densities. Note that the linear term for observable $\hat A$ is independent of the spin [see Eq.~\eqref{eq:linear-coeff-A}], so all spin densities (vanishing and nonvanishing) have the same slope $1/(10\sqrt{3})$ in the thermodynamic limit. This is also the slope one obtains using Eq.~\eqref{eq:linear-coeff} when carrying out the averages over the entire $2^L$-dimensional Hilbert space. On the other hand, observable $\hat B$ has a slope that vanishes in the thermodynamic limit for vanishing spin densities and approaches an $O(1)$ number for nonzero spin densities [see Eq.~\eqref{eq:linear-coeff-B}]. 

Since $\hat B$ (of the form $\hat T^{(2)}_0$) is a spherical tensor operator that is not rotationally invariant and $\hat H$ (of the form $\hat T^{(0)}_0$) is rotationally invariant, the product $\hat B\hat H^n$ is also a $\hat T^{(2)}_0$ operator for all $n$. Therefore, the infinite temperature expectation value $\braket{\hat B\hat H^n}_S$ computed in a sector with fixed total spin $S$ but all possible magnetization values $M=-S,\ldots,S$ vanishes as a consequence of the Wigner-Eckart theorem~\eqref{eq:Wigner-EckartTheorem}:
\begin{align}\label{eq:<BH^n>}
    \braket{\hat B\hat H^n}_S&=\frac{1}{2S+1}\sum_{M=-S}^S \braket{\hat B\hat H^n}_{S,M}\\&=\frac{\braket{\hat B\hat H^n}_{S,M=0}}{\braket{S0|S0;2,0}}\frac{\sum_{M=-S}^S \braket{SM|SM;2,0}}{2S+1}=0\,,\nonumber
\end{align}
where we used that the sum of Clebsch-Gordan coefficients $\sum_{M=-S}^S \braket{SM|SM;r0}=(2S+1)\delta_{r,0}=0$ for $r=2$~\cite{clebsch_gordan_formula_book_1988}. Therefore, although the diagonal function for $\hat B$ can exhibit an energy dependence within sectors with fixed total spin $S$ and magnetization $M=0$ [the leading-order linear dependence is reported in Eq.~\eqref{eq:linear-coeff-B}], it becomes structureless, \ie $B(E_\alpha,S)=0$ after including all magnetization sectors. This occurs because $\hat B$ does not overlap with any power of the Hamiltonian, \ie $\braket{\hat B\hat H^n}_S=0$ [see Eq.~\eqref{eq:<BH^n>}], and also applies to the average over the entire $2^L$-dimensional Hilbert space. Similar behaviors of matrix elements of observables in local Hamiltonians have been observed and discussed in the absence of non-Abelian symmetries~\cite{mierzejewski_vidmar_20, schonle_autocorrelation_2021}.  

To quantify the scaling of the fluctuations of $O_{\alpha\alpha}$ about the smooth function $O(E_\alpha,S_\alpha)$ with increasing system size, we calculate the average
\begin{equation}\label{eq: diagonal fluctuations}
    \langle \delta O_{\alpha\alpha}\rangle =\langle|O_{\alpha\alpha}-\overline{O_{\alpha\alpha}}| \rangle
\end{equation}
of the fluctuations of $O_{\alpha\alpha}$ about the mean value $\overline {O_{\alpha\alpha}}$ [computed as a running average $\overline{(\ldots)}$ over 50 states centered at $E_\alpha$]. The average $\langle \delta O_{\alpha\alpha} \rangle$ is calculated in the central 50\% of the energy spectrum. Since the observables considered here are (translationally invariant) intensive sums of local operators, which would need to be multiplied by $L^{1/2}$ to be properly normalized~\cite{mierzejewski_vidmar_20, patrycja_rafal_24}, the diagonal part of the ETH ansatz in Eq.~\eqref{eq:Fullmatrixelementnon-AbelianETH} needs to be modified as done for the traditional ETH~\cite{leblond_2019}
\begin{align}
    O_{\alpha\alpha}&=O(E_\alpha,S_\alpha)\nonumber \\
    &\qquad + \frac{e^{-S_{\text{th}}(E_\alpha,S_\alpha)/2}}{\sqrt{L}}f_O(E_\alpha,0;S_\alpha,0)R_{\alpha\alpha}\,.\label{eq:adjusteddiagonalnon-AbelianETH}
\end{align}
Consequently, at the center of the energy spectrum, where $e^{S_{\text{th}}(E_\alpha,S_\alpha)}\simeq D$ (where $D$ is the Hilbert-space dimension of the specific symmetry sector considered), we expect the average fluctuations $\langle \delta O_{\alpha\alpha}\rangle$ in a system that exhibits eigenstate thermalization to scale as $(LD)^{-1/2}$.

In Figs.~\ref{fig:DiagonalvsEnergydensity}(b) and~\ref{fig:DiagonalvsEnergydensity}(d), we show the average fluctuations of the observables in the left panels. In the nonintegrable regime (main panels), the average fluctuations vanish exponentially as $(LD)^{-\gamma}$ with $\gamma\approx 0.5$ (see the straight lines, which are obtained fitting $\gamma$ from the data). We note that in our fits $\gamma$ is closer to $0.5$ for the cases with $s\simeq 0$ than for $s=1/4$. This is expected as the former have larger Hilbert space dimensions $D$ and therefore suffer from weaker finite-size effects. In the integrable regime, on the other hand, the average fluctuations [inset in Fig.~\ref{fig:DiagonalvsEnergydensity}(d)] exhibit a slower algebraic decay $L^{-\delta}$ with $\delta \approx 1$. This exponent complies with the requirement that, for the (translationally invariant) intensive observables considered here (no-matter the model), the fluctuations over the entire Hilbert space must decay at least as fast as $L^{-1/2}$~\cite{patrycja_rafal_24}.

\begin{figure}[!t]
    \includegraphics[width=0.90\columnwidth]{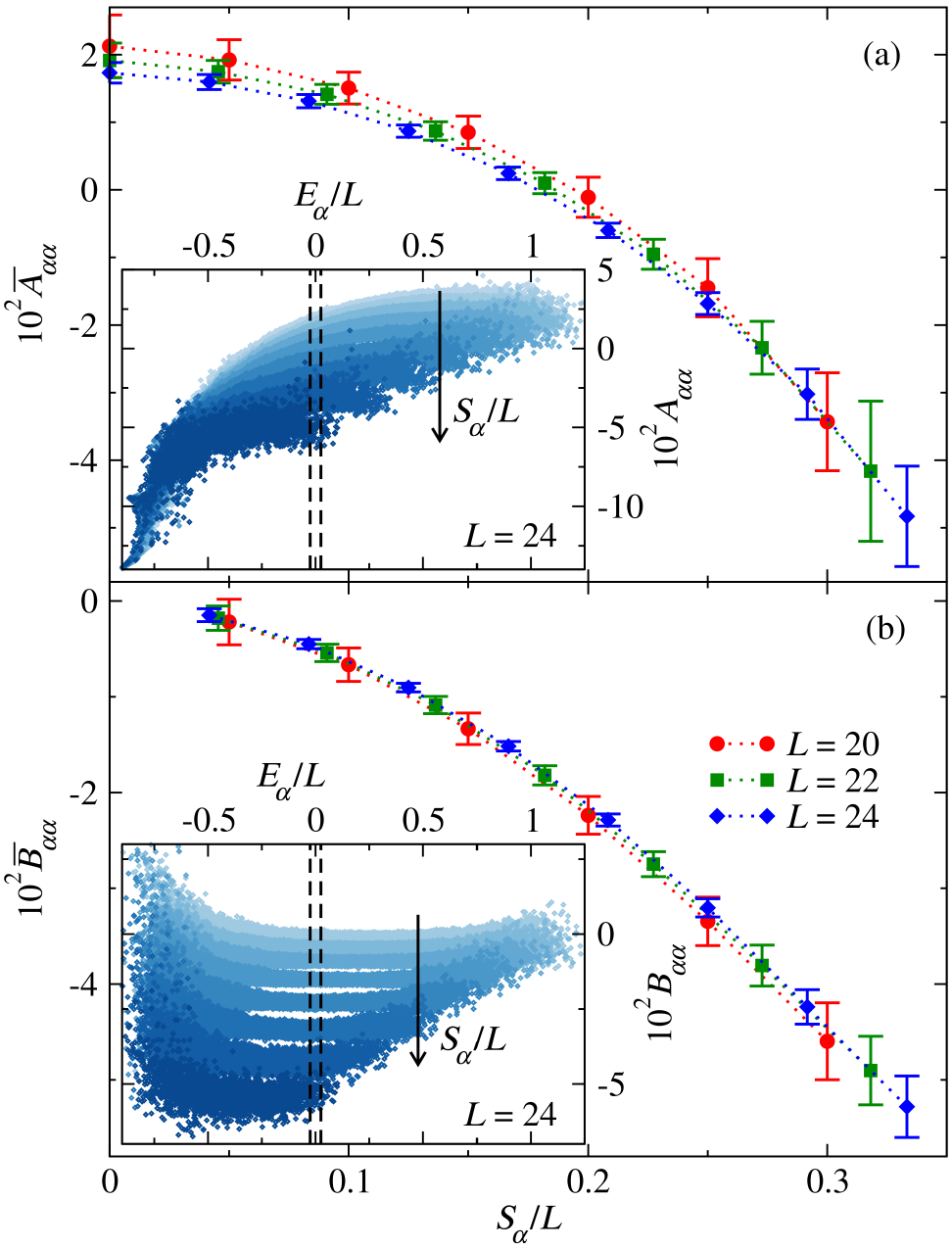}
    \vspace{-0.2cm}
    \caption{Diagonal matrix elements for $\hat A$ (a)  and $\hat B$ (b) in the nonintegrable Hamiltonian eigenstates ($\lambda=3$), averaged over a narrow window in energy density $|E_\alpha|/L\leqslant 0.025$, vs the spin density $S_\alpha/L$. The error bars correspond to the standard deviation representing the fluctuations about the mean value. The inset shows the diagonal matrix elements vs energy density at $L=24$ for spins $S=0,1,\ldots,8$ (a) and $S=1,2,\ldots,8$ (b), where the darker shades represent larger spin (spin density) values as indicated by the arrow. The dashed lines depict the fixed energy density window $|E_\alpha|/L\leqslant 0.025$ used to obtain the averages in the main panel.}
    \label{fig:DiagonalvsSpindensity}
\end{figure}

The results reported in Fig.~\ref{fig:DiagonalvsEnergydensity} probe one aspect of the diagonal part of the ETH ansatz in Eq.~\eqref{eq:Fullmatrixelementnon-AbelianETH}, namely, the smoothness of $O(E_\alpha,S_\alpha)$ vs the energy density $E_\alpha/L$ at different spin densities $S_\alpha/L$. The smooth energy dependence of the diagonal part of the ETH has been extensively studied in models with no continuous symmetries and with Abelian U(1) symmetry at fixed magnetization (particle number)~\cite{rigol_2008, rigol_09a,*rigol_09b, kim_testing_ETH_2014, beugeling_2014,*beugeling_offdiag_2015, mondaini_2016, mondaini_2017, jansen_2019, leblond_2019, leblond_2020, schonle_autocorrelation_2021, wang_2024}. Next, we study the smoothness of $O(E_\alpha,S_\alpha)$ vs the spin density $S_\alpha/L$ at fixed energy density $E_\alpha/L$.

The insets in Figs.~\ref{fig:DiagonalvsSpindensity}(a) and~\ref{fig:DiagonalvsSpindensity}(b) show the diagonal matrix elements of $\hat A$ and $\hat B$, respectively, plotted vs $E_\alpha/L$ for a fixed system size ($L=24$) and different values of $S_\alpha$ (a darker shade is used for larger $S_\alpha$) in the nonintegrable regime (see also Ref.~\cite{noh_2023,lasek_24}). Those results indicate that, at fixed $E_\alpha$ and $L$, the matrix elements of both observables are smooth functions of $S_\alpha$. To further explore this dependence, in the main panels in Figs.~\ref{fig:DiagonalvsSpindensity}(a) and~\ref{fig:DiagonalvsSpindensity}(b) we plot $A_{\alpha\alpha}$ and $B_{\alpha\alpha}$, respectively, averaged over a narrow window in energy density $|E_\alpha|/L\leqslant 0.025$ (highlighted in the inset), vs the spin density. We show results for the largest three system sizes considered. The smoothness of the curves and their collapse for different system sizes suggest that the non-Abelian ETH describes the spin dependence of the diagonal matrix elements of observables in nonintegrable models.

\section{Off-diagonal matrix elements}
\label{sec:Off-diag}

In this section, we study the off-diagonal matrix elements $O_{\alpha\beta}$ for the two observables $\hat O=\hat A,\hat B$ and probe various aspects of the off-diagonal  part of the non-Abelian ETH. For our intensive translationally invariant observables, because of their normalization (as mentioned in Sec.~\ref{sec: Diag}), the off-diagonal part of Eq.~\eqref{eq:Fullmatrixelementnon-AbelianETH} needs to be modified to read
\begin{equation}
    O_{\alpha\beta} = \frac{e^{-S_{\text{th}}(\bar E,\bar S)/2}}{\sqrt{L}}f_O(\bar E,\omega;\bar S,\nu)R_{\alpha\beta}\,.\label{eq:adjustedoff-diagonalnon-AbelianETH}
\end{equation}

The off-diagonal ETH has three main components, some of which are common to the fluctuations associated with the diagonal ETH discussed in the previous section. The (to a good approximation) Gaussian random variable $R_{\alpha\beta}$ with zero mean and unit variance, the exponential (in the system size) suppression factor $e^{-S_{\text{th}}(\bar E,\bar S)/2}/\sqrt{L}$, and the smooth function $f_O(\bar E,\omega; \bar S,\nu)$. In what follows, we probe them independently, as done in the past for systems with Abelian symmetries (see, e.g., Refs.~\cite{leblond_2019, leblond_2020}). 

We compute the off-diagonal matrix elements $O_{\alpha\beta}$ between energy eigenstates for which $\bar E=(E_\alpha+E_\beta)/2$ lies at the center of the energy spectrum. Specifically, we take $|\bar E - E_0|/L\leqslant 0.025$, where $E_0=\mathrm{Tr}(\hat H)/D$. Observable $\hat A$ only has nonvanishing off-diagonal matrix elements between energy eigenstates with the same spin ($S_\alpha=S_\beta$). Observable $\hat B$ has nonvanishing off-diagonal matrix elements both between energy eigenstates with the same spin ($S_\alpha=S_\beta$) and different spins ($S_\alpha\neq S_\beta$). In Sec.~\ref{sec:Off-diag-equalspins} (Sec.~\ref{sec:Off-diag-unequalspins}), we study the off-diagonal matrix elements of $\hat A$ and $\hat B$ ($\hat B$) between energy eigenstates with the same spin (different spins).

\subsection{Same spin} 
\label{sec:Off-diag-equalspins}

We first probe the (close to) Gaussian nature~\footnote{The fact that correlations must exist between matrix elements of observables (namely, that their fluctuations cannot be completely random) becomes apparent when studying two-observable correlation functions~\cite{dalessio_quantum_2016}, and it has been a topic of much interest in the context of out-of-time-order correlators~\cite{foini_kurchan_19, chan_deluca_19, murthy_srednicki_19, brenes_pappalardi_21, wang_lamann_22}.} of the fluctuations of the off-diagonal elements associated with $R_{\alpha\beta}$ in Eq.~\eqref{eq:adjustedoff-diagonalnon-AbelianETH}. For that, as done in earlier works in systems with Abelian U(1) symmetry~\cite{leblond_2019, leblond_2020}, we calculate
\begin{equation}\label{eq:Gaussianitymeasure}
    \Gamma_O(\omega)=\frac{\overline{|O_{\alpha\beta}|^2}}{\overline{|O_{\alpha\beta}|}^2}\,,
\end{equation}
where $\overline{(\ldots)}$ is a running average computed within $\omega$-windows of small size $\Delta\omega$ (we take $\Delta\omega=0.175$) for values of $\omega$ separated by $\delta\omega$ (we take $\delta\omega=0.025$). If $O_{\alpha\beta}$ corresponds to a Gaussian random variable with zero mean, then $\Gamma_O(\omega)$ must be equal to $\pi/2$. In other words, the proximity of $\Gamma_O(\omega)$ to $\pi/2$ quantifies the proximity of $O_{\alpha\beta}$ to a Gaussian distribution. 

\begin{figure}[!t]
    \includegraphics[width=0.98\columnwidth]{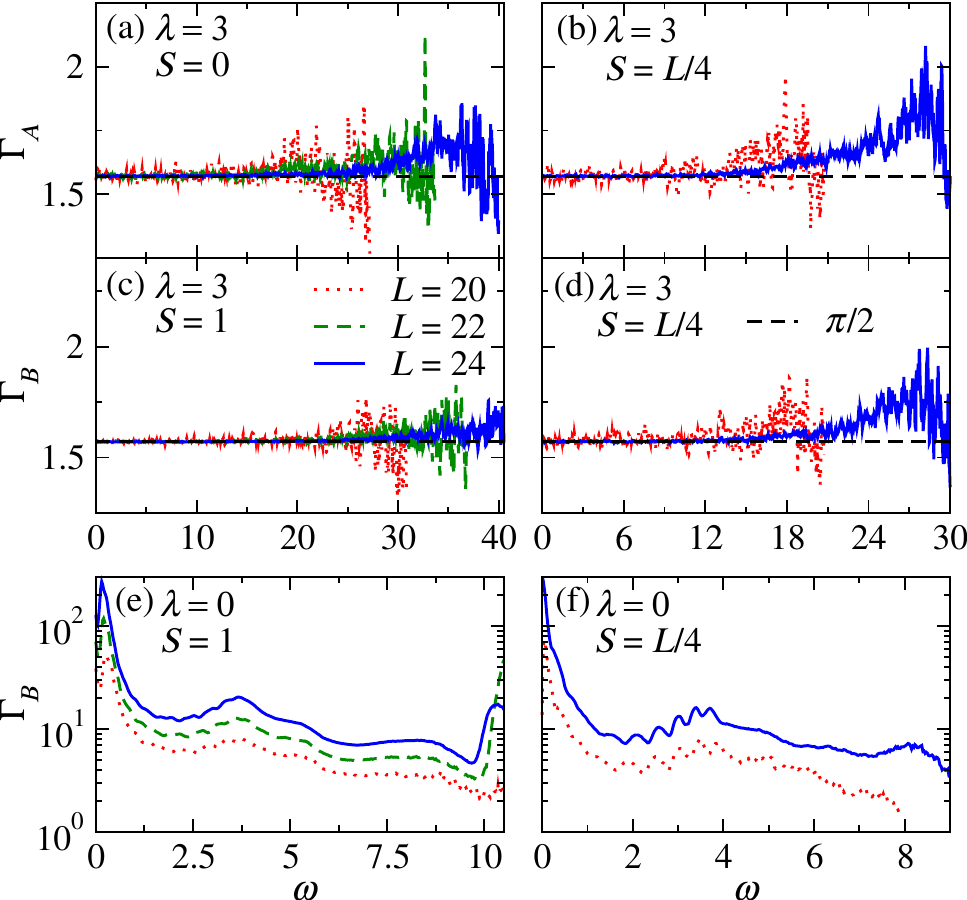}
    \vspace{-0.1cm}
    \caption{$\Gamma_O$ [see Eq.~\eqref{eq:Gaussianitymeasure}] vs $\omega$ for: (a)--(d) $\hat O = \hat A$ and $\hat O =\hat B$ in the nonintegrable regime ($\lambda=3$), and (e),(f) $\hat B$ at integrability ($\lambda=0$). For $\hat A$ we show results at (a) $S=0$ and (b) $S=L/4$, while for $\hat B$ we show results at (c),(e) $S=1$ and (d),(f) $S=L/4$. The horizontal dashed lines in (a)--(d) mark $\Gamma=\pi/2$, corresponding to a Gaussian distribution. For both observables, $\overline{|O_{\alpha\beta}|^2}$ and $\overline{|O_{\alpha\beta}|}^2$ in $\Gamma_O$ were computed using energy eigenstates with $|\bar E-E_0|/L\leqslant 0.025$, as a running average over $\omega$-windows of size $\Delta\omega=0.175$ for values of $\omega$ separated by $\delta\omega=0.025$.}
    \label{fig:Gaussianity}
\end{figure} 

Figure~\ref{fig:Gaussianity} shows the results for $\Gamma_O$ computed for $\hat O=\hat A$ [Figs.~\ref{fig:Gaussianity}(a) and~\ref{fig:Gaussianity}(b)] and $\hat O=\hat B$ [Figs.~\ref{fig:Gaussianity}(c) and~\ref{fig:Gaussianity}(d)] for the nonintegrable regime in chains with $L=20$ -- 24 sites. For $\hat A$ we show results at $S=0$ (a) and $S=L/4$ (b), while for $\hat B$ we show results at $S=1$ (c) and $S=L/4$ (d). The black dashed lines indicate $\Gamma_O=\pi/2$, corresponding to a Gaussian distribution. For all the results shown in Figs.~\ref{fig:Gaussianity}(a)--\ref{fig:Gaussianity}(d), $\Gamma_O$ remains close to $\pi/2$ for $\omega\lesssim 20$ ($\omega\lesssim 12$) for $S=0,1$ ($S=L/4$) indicating that the off-diagonal matrix elements exhibit a nearly Gaussian distribution. The deviations from $\pi/2$ observed for larger values of $\omega$ are a consequence of finite-size effects. They occur because those are matrix elements between energy eigenstates that are close to the edges of the energy spectrum. (Notice that the window of values of $\omega$ for which $\Gamma_O\approx \pi/2$ increases as $L$ increases and the spectrum becomes wider). The results for $\Gamma_B$ at integrability, shown in Figs.~\ref{fig:Gaussianity}(e) and~\ref{fig:Gaussianity}(f), are on the other hand strongly dependent on the value of $\omega$ and the systems-size (notice the log scale in the $y$ axes). This makes apparent the non-Gaussian nature of $O_{\alpha\beta}$ at integrability. 

\begin{figure}[!t]
    \includegraphics[width=0.98\columnwidth]{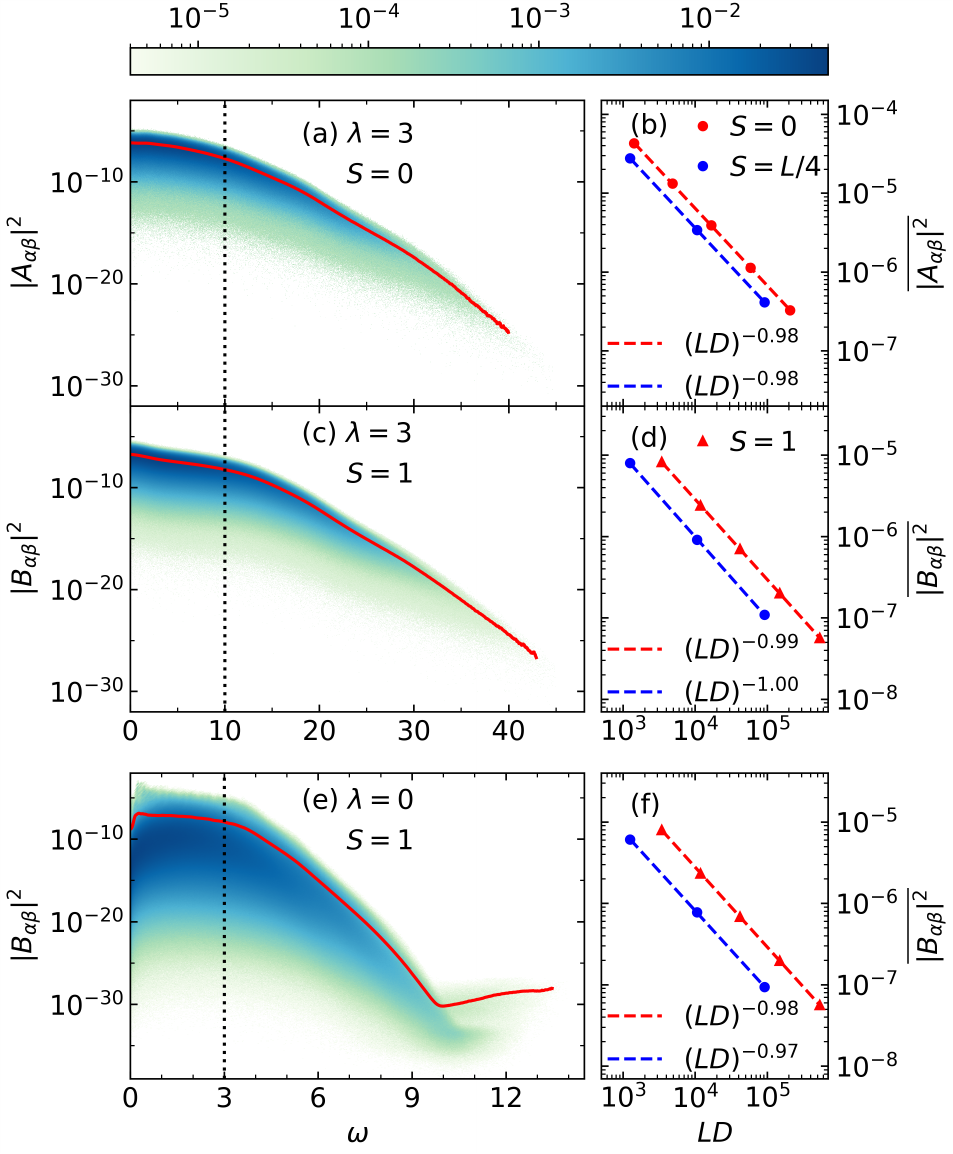}
    \vspace{-0.2cm}
    \caption{Left column: Color coded normalized histogram of the distribution of $|O_{\alpha\beta}|^2$ vs $\omega=E_\alpha-E_\beta$ in a chain with $L=24$ sites for energy eigenstates with $|\bar E-E_0|/L\leqslant 0.025$. (a) $\hat O=\hat A$ ($S=0$) and (c) $\hat O=\hat B$ ($S=1$) in the nonintegrable regime ($\lambda=3$), and (e) $\hat O=\hat B$ ($S=1$) at integrability ($\lambda=0$). The solid red curves correspond to the variance $\overline{|O_{\alpha\beta}|^2}$ computed as a running average with a sliding window of size $\Delta\omega=0.175$ centered at values of $\omega$ separated by $\delta\omega=0.025$. Right column: The variance $\overline{|O_{\alpha\beta}|^2}$ vs $LD$ obtained by averaging $|O_{\alpha\beta}|^2$ over values of $\omega$ up to the black dotted line in the left column for $S=0,\,1$ ($L=16,\,18,\,\ldots,\, 24$) and $S=L/4$ ($L=16,\,20,\, 24$, histogram not shown). The dashed lines are power law fits $\propto (LD)^\gamma$, with the fitting parameter $\gamma$ (reported in the legends) found to be close to $-1$ both in the nonintegrable (b),(d) and integrable (f) regimes.}  
    \label{fig:2DHistogram}
\end{figure}

Next, we study the distribution of $|O_{\alpha\beta}|^2$ vs $\omega=E_\alpha-E_\beta$. In the left column of Fig.~\ref{fig:2DHistogram}, we plot the normalized 2D histograms for $|O_{\alpha\beta}|^2$ vs $\omega$ for the observables $\hat A$ ($S=0$) and $\hat B$ ($S=1$) in a chain of size $L=24$ (qualitatively similar results were obtained for $S=L/4$, not shown). At fixed $\omega$, the distributions of $|O_{\alpha\beta}|^2$ in the nonintegrable regime [Figs.~\ref{fig:2DHistogram}(a) and~\ref{fig:2DHistogram}(c)] are narrow, which contrasts the broad distribution in the integrable regime [Fig.~\ref{fig:2DHistogram}(e)]. The values of $|O_{\alpha\beta}|^2$ exhibit an initial slow decay to intermediate values of $\omega$ (roughly up to the black dotted line) followed by a fast decay for higher frequencies. The latter decay is better seen in the variance $\overline{|O_{\alpha\beta}|^2}$ as a function of $\omega$ (solid red line), which is calculated as a running average over the values of $|O_{\alpha\beta}|^2$ within $\omega$-windows of size $\Delta \omega=0.175$ for values of $\omega$ separated by $\delta\omega=0.025$.

The results in the right column in Fig.~\ref{fig:2DHistogram} probe the expected exponential decay $e^{-S_{\text{th}}(\bar E,\bar S)/2}/\sqrt{L}$ of the off-diagonal matrix elements advanced by Eq.~\eqref{eq:adjustedoff-diagonalnon-AbelianETH}. Since we consider energy eigenstates belonging to the center of the spectrum with $\bar E$ given by $|\bar E - E_0|/L\leqslant 0.025$, the exponential factor roughly equals the Hilbert space dimension of the symmetry sector considered, namely, $e^{S_{\text{th}}(\bar E,\bar S)}\approx D$. Therefore, the variance $\overline{|O_{\alpha\beta}|^2}$ is expected to exhibit an overall $1/LD$ scaling. In the right column of Fig.~\ref{fig:2DHistogram}, we plot the variance $\overline{|O_{\alpha\beta}|^2}$ vs $LD$, where $\overline{|O_{\alpha\beta}|^2}$ is calculated over the matrix elements with $\omega<10$ ($\omega<3$) in the nonintegrable (integrable) regime indicated by the black dotted line in the left column of Fig.~\ref{fig:2DHistogram}. The variance for the observables $\hat A$ [Fig.~\ref{fig:2DHistogram}(b)] and $\hat B$ [Fig.~\ref{fig:2DHistogram}(d)] in the nonintegrable regime is consistent with the expected $\sim (LD)^{-1}$ decay. Remarkably, the variance for observable $\hat B$ in the integrable regime [Fig.~\ref{fig:2DHistogram}(f)] shows the same $(LD)^{-1}$ scaling. This has also been observed in integrable systems with Abelian U(1) symmetry~\cite{leblond_2019, leblond_2020}.

Next, we explore the $\omega$ dependence of the spectral function $|f_O(\bar E,\omega; \bar S,\nu)|^2$ [see Eq.~\eqref{eq:adjustedoff-diagonalnon-AbelianETH}] corresponding to the variance of the off-diagonal matrix elements $\overline{|O_{\alpha\beta}|^2}$ for the energy eigenstates in the center of the spectrum (with $\bar E$ satisfying $|\bar E - E_0|/L\leqslant 0.025$). (In this section, the energy eigenstates have the same spin $S$ so $\bar S=S$ and $\nu=S_\alpha-S_\beta=0$.) From the ETH ansatz, the spectral function for an observable $\hat O$ in the nonintegrable regime can be obtained from the matrix elements as
\begin{equation}\label{eq:spectralfunction}
    |f_O(\bar E\approx E_0,\omega;\bar S,\nu)|^2= LD\,\overline{|O_{\alpha\beta}|^2}\,.
\end{equation}
In the integrable regime, $|f_O(\bar E\!\approx\! E_0,\omega;\bar S,\nu)|^2$ as defined above is still a well-defined smooth function even though the matrix elements are not described by ETH~\cite{leblond_2019, leblond_2020}.    

\begin{figure}[!t]
    \includegraphics[width=0.98\columnwidth]{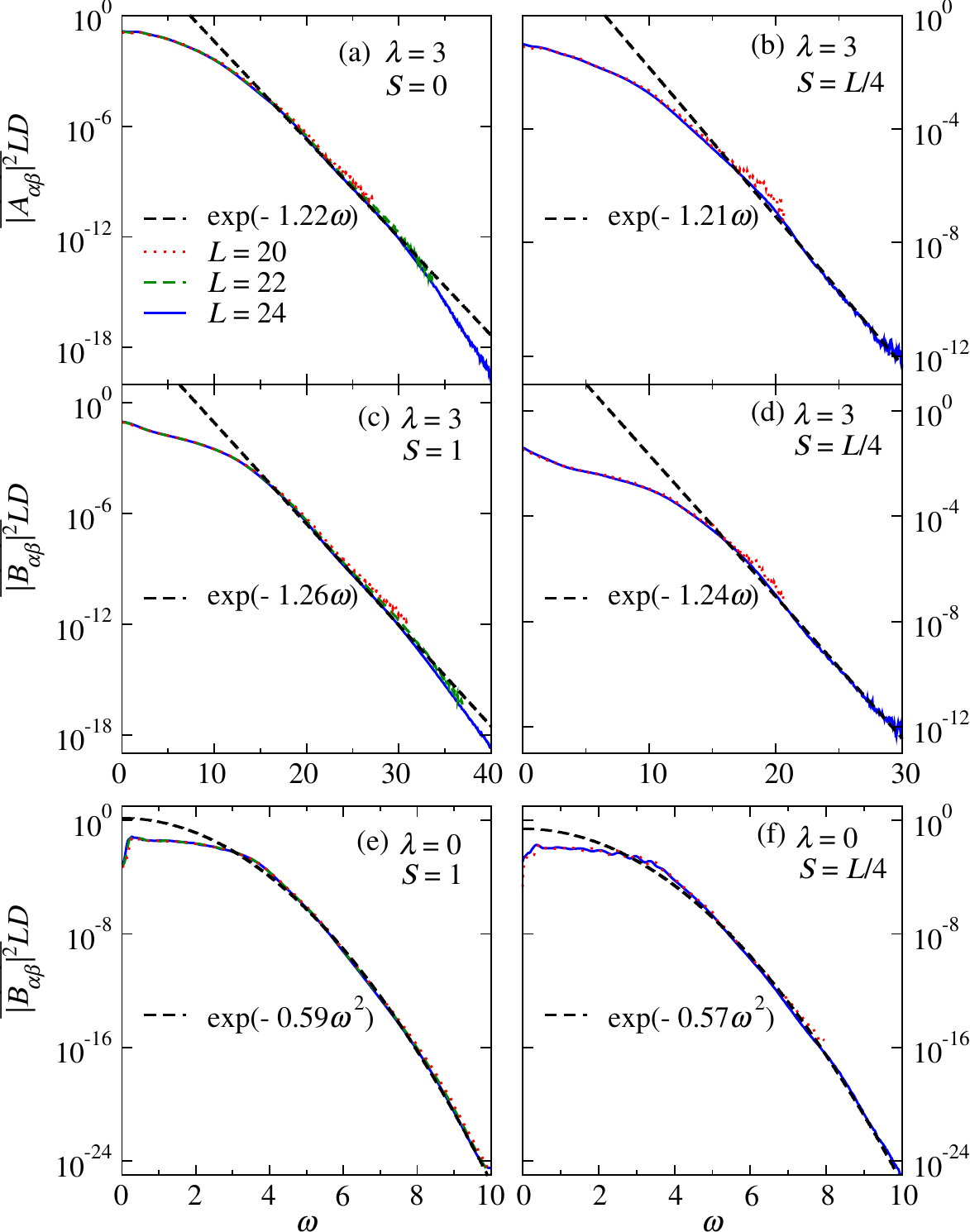}
    \vspace{-0.2cm}
    \caption{Spectral functions vs $\omega$ for: (a)--(d) $\hat A$ and $\hat B$ in the nonintegrable regime ($\lambda=3$), and (e),(f) $\hat B$ at integrability ($\lambda=0$). For $\hat A$ we show results at (a) $S=0$ and (b) $S=L/4$, while for $\hat B$ we show results at (c),(e) $S=1$ and (d),(f) $S=L/4$. The black dashed lines correspond to exponential [Gaussian] fits, $\propto \exp(-a\omega)$ [$\propto \exp(-b\omega^2)$] for chains with $L=24$ sites in the nonintegrable [integrable] regime.}
    \label{fig:highfrequencyspectralfunction}
\end{figure} 

In Fig.~\ref{fig:highfrequencyspectralfunction}, we plot the results for the spectral function as defined in Eq.~\eqref{eq:spectralfunction} for the two observables $\hat A$ and $\hat B$ in chains with $L=20$ -- 24 sites. The curves collapse as expected. After the previously mentioned initial slow observable-dependent decay, the fast decay of the spectral function is consistent with being exponential up to $\omega \approx 30$ in the nonintegrable regime [see the black dashed lines in Figs.~\ref{fig:highfrequencyspectralfunction}(a)--\ref{fig:highfrequencyspectralfunction}(d)]. For $S=0$ and 1 in Figs.~\ref{fig:highfrequencyspectralfunction}(a) and~\ref{fig:highfrequencyspectralfunction}(c), respectively, the spectral function decays faster than exponentially for $\omega \gtrsim 30$. At integrability, on the other hand, the spectral function appears to be Gaussian right after the initial slow decay [see Figs.~\ref{fig:highfrequencyspectralfunction}(e) and~\ref{fig:highfrequencyspectralfunction}(f)]. Such behaviors of the spectral function have also been observed in systems with Abelian symmetries~\cite{leblond_2019, jansen_2019, leblond_2020, zhang_22}.

\begin{figure}[!t]
    \includegraphics[width=0.98\columnwidth]{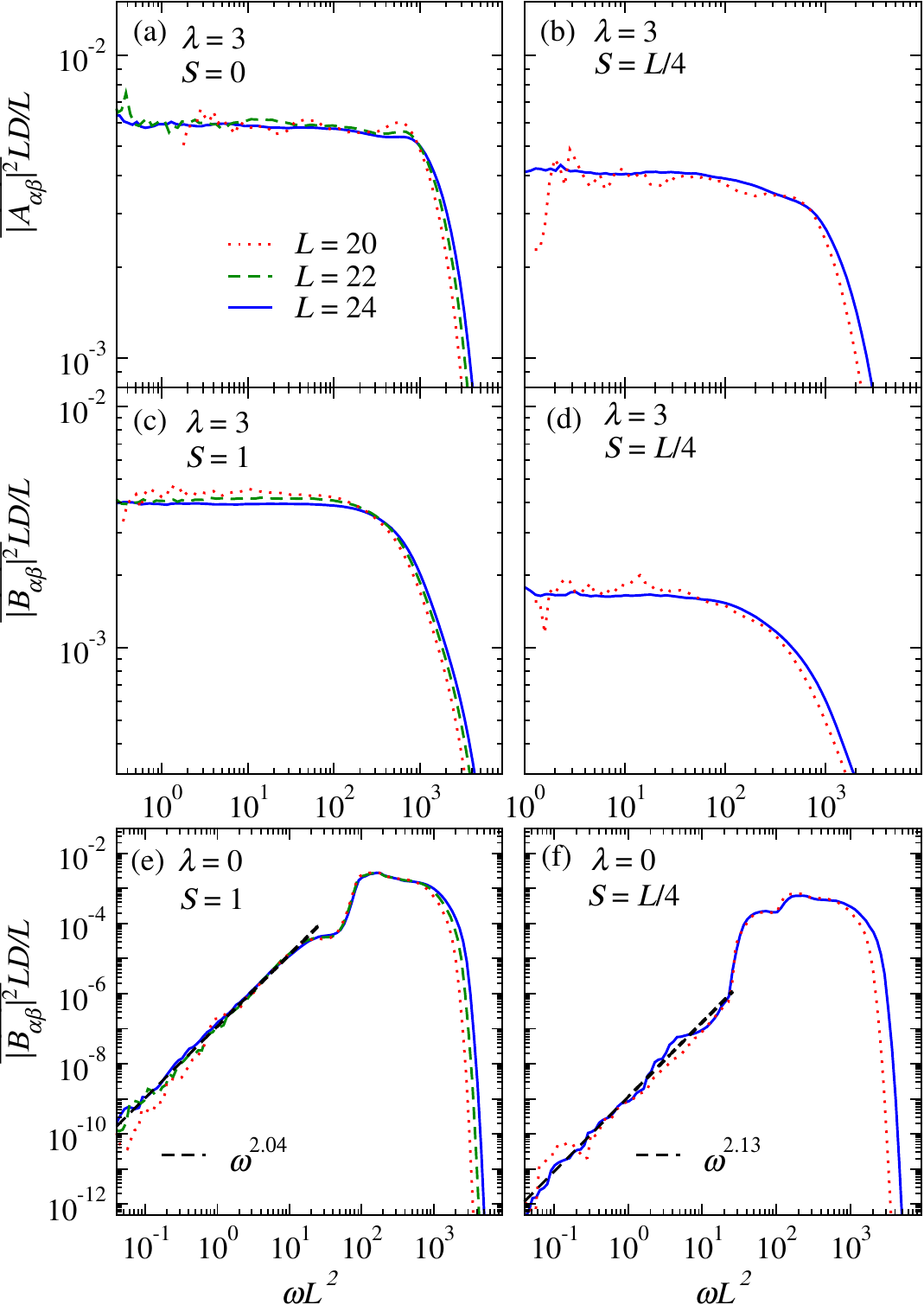}
    \vspace{-0.2cm}
    \caption{Low-frequency behavior of the spectral functions shown in Fig.~\ref{fig:highfrequencyspectralfunction} plotted vs $\omega L^2$. The black dashed lines in (e) and (f) correspond to power law fits $\propto \omega^a$ for chains with $L=24$ sites, with the fitted value of $a$ (reported in the legends) found to be $a\approx 2$.} 
    \label{fig:lowfrequencyspectralfunction}
\end{figure} 

In Fig.~\ref{fig:lowfrequencyspectralfunction}, we plot the corresponding results for the spectral function in the low-frequency regime. As $\omega\rightarrow0$, below the Thouless energy $\omega_T\propto 1/L^2$, one expects random matrix behavior to set in for nonintegrable systems and the spectral function to become frequency independent~\cite{dalessio_quantum_2016}. This is indeed observed in the plots in Figs.~\ref{fig:lowfrequencyspectralfunction}(a)--\ref{fig:lowfrequencyspectralfunction}(d). However, how the value of the spectral function in the plateau scales with the system size depends on the observable considered~\cite{schonle_autocorrelation_2021}. For $\hat A$, which overlaps with the Hamiltonian [see Eq.~\eqref{eq:linear-coeff-A}], we expect $LD\,\overline{|A_{\alpha\beta}|^2}$ to be proportional to $L$ due to diffusion~\cite{dalessio_quantum_2016}. Indeed, in Figs.~\ref{fig:lowfrequencyspectralfunction}(a) and~\ref{fig:lowfrequencyspectralfunction}(b) the results collapse at low frequency when plotting $LD\,\overline{|A_{\alpha\beta}|^2}/L$ vs $\omega L^2$. For $\hat B$ when averaging over all possible magnetization sectors at fixed spin, since the overlap of $\hat B$ with all powers of $\hat H$ vanishes [see Eq.~\eqref{eq:<BH^n>}], we expect $LD\,\overline{|B_{\alpha\beta}|^2}$ to be system size independent as predicted by random-matrix theory. However, when averaging at fixed magnetization and spin in finite systems (which is the case here), since $\hat B$ overlaps with $\hat H$ [see Eq.~\eqref{eq:linear-coeff-B}], we expect $LD\,\overline{|B_{\alpha\beta}|^2}$ to depend on $L$ [see Figs.~\ref{fig:lowfrequencyspectralfunction}(c) and~\ref{fig:lowfrequencyspectralfunction}(d)]. Only for vanishing spin density in the thermodynamic limit do we expect $LD\,\overline{|B_{\alpha\beta}|^2}$ to be system-size independent. The results in Fig.~\ref{fig:lowfrequencyspectralfunction}(c) are consistent with that expectation, as the dependence of $LD\,\overline{|B_{\alpha\beta}|^2}$ on $L$ slows down with increasing system size (notice that dividing $LD\,\overline{|B_{\alpha\beta}|^2}$ by $L$ results in a plateau that moves down with increasing $L$).

In contrast to the behaviors in the quantum-chaotic regime, the low-frequency spectral function for observable $\hat B$ in the integrable regime [see Figs.~\ref{fig:lowfrequencyspectralfunction}(e) and~\ref{fig:lowfrequencyspectralfunction}(f)] is $\propto \omega^2$, namely, it vanishes as $\omega\rightarrow 0$ as expected for observables that preserve integrability when added as perturbations to the Hamiltonian~\cite{pandey_claeys_20, leblond_2020, leblond_sels_21, kim_polkovnikov_24}. All the results discussed in this section are qualitatively similar to those obtained for systems with Abelian U(1) symmetry. The SU(2) symmetry does not qualitatively change eigenstate thermalization for sectors with fixed spin. 

\subsection{Different spins}
\label{sec:Off-diag-unequalspins}

Next, we study the off-diagonal matrix elements $B_{\alpha\beta}$ corresponding to energy eigenstates belonging to different spin sectors $S_\alpha\neq S_\beta$, with average energies in the center of the spectrum, $|\bar E-E_0|/L\leqslant 0.025$. The observable $\hat B$ is a spherical tensor operator of the form $\hat T^{(2)}_0$ \ie it can connect energy eigenstates $\ket{E_\alpha S_\alpha M_\alpha}$ and $\ket{E_\beta S_\beta M_\beta}$ with the same magnetization $M_\alpha=M_\beta$ and with spins $|S_\beta-2|\leqslant S_\alpha \leqslant S_\beta+2$. However, within the zero magnetization sector, the spin-inversion ($Z_2$) symmetry imposes an additional constraint that disallows the combination $S_\alpha=|S_\beta\pm1|$, since $Z_2$ partitions the even and odd spin states into sectors with different parities. Thus, $\hat B$ can connect states with $S_\alpha=S_\beta$ (as studied in Sec.~\ref{sec:Off-diag-equalspins}), or $S_\alpha=|S_\beta\pm2|$, which we study next. Within a simultaneous symmetry sector, the Hilbert space dimension corresponding to spin $S_\alpha$, say $D_\alpha$, is in general different from the dimension $D_\beta$ for spin $S_\beta \neq S_\alpha$. We take $D\equiv\sqrt{D_\alpha D_\beta}$ to be the ``dimension'' of the rectangular $D_\alpha \times D_\beta$ block corresponding to the symmetry sectors with spin combination $(S_\alpha,S_\beta)$. 

\begin{figure}[!t]
    \includegraphics[width=0.98\columnwidth]{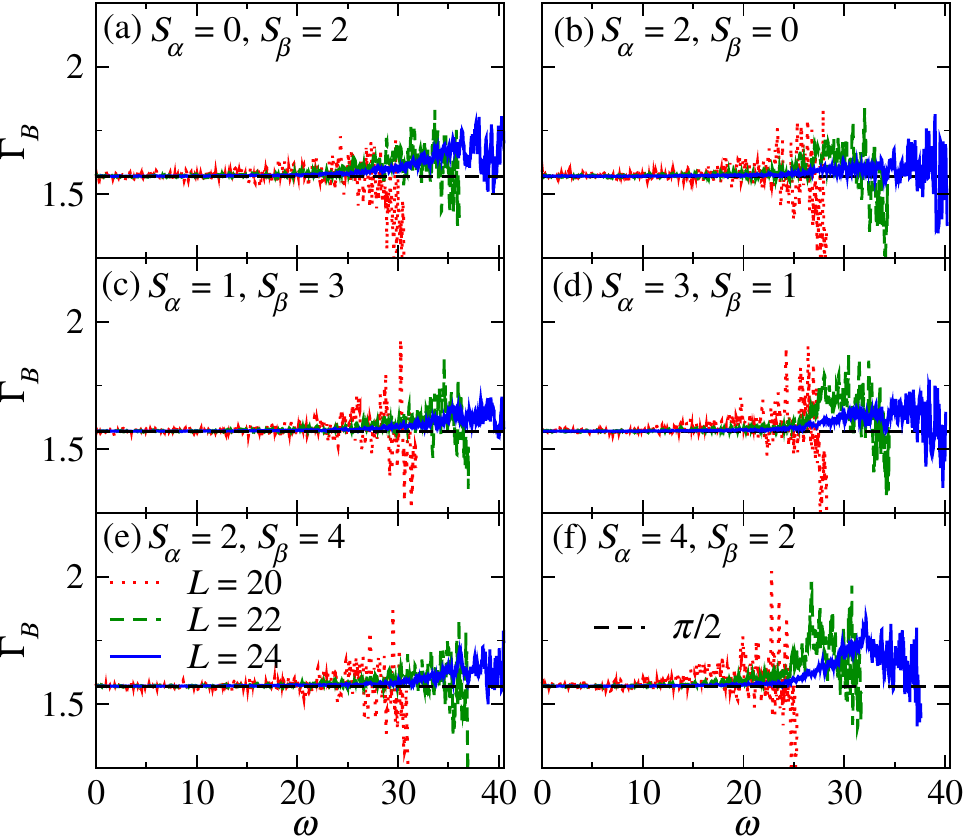}
    \vspace{-0.1cm}
    \caption{$\Gamma_B$ [see Eq.~\eqref{eq:Gaussianitymeasure}] vs $\omega$ for the nonintegrable ($\lambda=3$) regime in spin sectors labeled by $(S_\alpha,S_\beta)$ in chains with $L=20,\,22,\,24$. The left (right) column 
    show results for $\nu=S_\alpha-S_\beta=-2$ ($\nu=2$). The averages $\overline{|B_{\alpha\beta}|^2}$ and $\overline{|B_{\alpha\beta}|}^2$ used to calculate $\Gamma_B$ were computed using energy eigenstates with $|\bar E-E_0|/L\leqslant 0.025$, as a running average over $\omega$-windows of size $\Delta\omega=0.175$ for values of $\omega$ separated by $\delta\omega=0.025$.}
    \label{fig:differentspinsgaussianity}
\end{figure}

\begin{figure}[!b]
    \includegraphics[width=0.98\columnwidth]{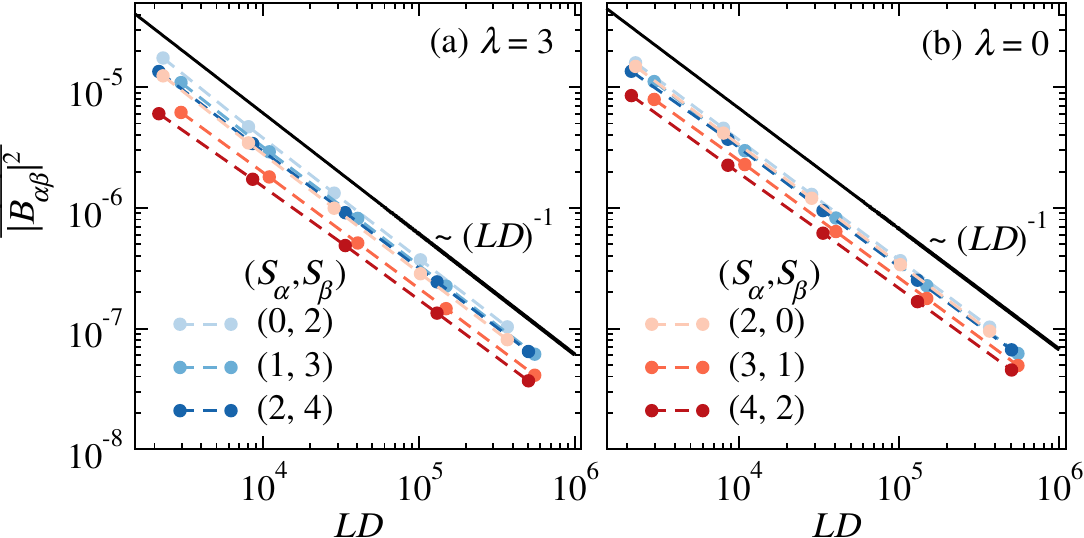}
    \vspace{-0.1cm}
    \caption{The variance $\overline{|B_{\alpha\beta}|^2}$ vs $LD$ for: (a) the nonintegrable ($\lambda=3$, averaged over $\omega\leq10$) and (b) the integrable ($\lambda=0$, averaged over $\omega\leq3$) regimes. We show results for the spin sectors $(S_\alpha,S_\beta)=(2,0),\,(3,1),\,(4,2)$ corresponding to a positive spin difference $\nu=S_\alpha-S_\beta=+2$, and for the spin sectors $(S_\alpha,S_\beta)=(0,2),\,(1,3),\,(2,4)$ corresponding to a negative spin difference of $\nu=-2$. The solid black lines are guides to the eye showing $(LD)^{-1}$ scalings.}
    \label{fig:differentspinsfluctuations}
\end{figure}
 
In Fig.~\ref{fig:differentspinsgaussianity}, we show results for $\Gamma_B$ [see Eq.~\eqref{eq:Gaussianitymeasure}] for different spin combinations $(S_\alpha,S_\beta)=(S,S+2)$ and $(S+2,S)$ with $S=0,\,1,\,2$ corresponding to negative (left column) and positive (right column) spin differences $\nu=S_\alpha-S_\beta=\pm2$, respectively. The results are qualitatively similar to those obtained for matrix elements between energy eigenstates with the same spin (see Fig.~\ref{fig:Gaussianity}). Therefore, the (close to) Gaussian nature of the fluctuations of the off-diagonal matrix elements of observables is not affected when energy eigenstates of different spin sectors are involved in the non-Abelian ETH.

We explore the scaling of the variance $\overline{|B_{\alpha\beta}|^2}$ with $LD$ in Fig.~\ref{fig:differentspinsfluctuations}. Like for the case with $S_\alpha=S_\beta$ (see the right column in Fig.~\ref{fig:2DHistogram}), for $S_\alpha\neq S_\beta$ we find that $\overline{|B_{\alpha\beta}|^2}$ exhibits a $\sim 1/LD$ decay in both the nonintegrable [Fig.~\ref{fig:differentspinsfluctuations}(a)] and integrable [Fig.~\ref{fig:differentspinsfluctuations}(b)] regimes, for sectors with positive and negative spin differences. This scaling in the nonintegrable regime is consistent with the off-diagonal non-Abelian ETH [see Eq.~\eqref{eq:adjustedoff-diagonalnon-AbelianETH}].  
         
\begin{figure}[!t]
    \includegraphics[width=0.98\columnwidth]{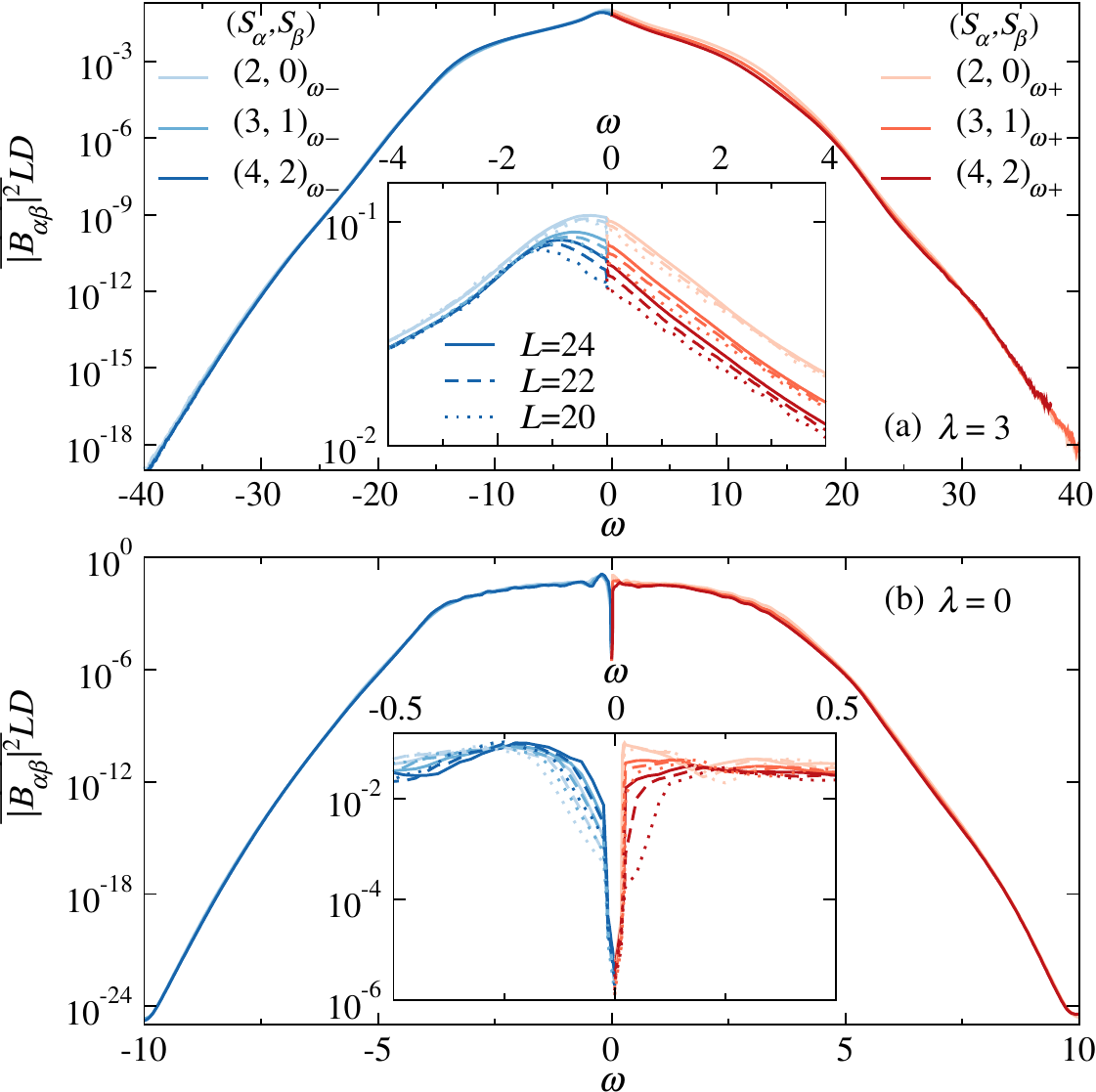}
    \vspace{-0.1cm}
    \caption{Spectral function of $\hat B$ vs $\omega$ between spin sectors $(S_\alpha,S_\beta)$ with positive spin differences $\nu=S_\alpha-S_\beta=+2$ in a chain with $L=24$ for: (a) the nonintegrable regime ($\lambda=3$), and (b) at integrability ($\lambda=0$). Darker shades in the plots correspond to larger spin averages $\bar S=(S_\alpha+S_\beta)/2$. The insets zoom into the region close to $\omega=0$, and include results for different system sizes: $L=24$ (solid lines), $L=22$ (dashed lines), and $L=20$ (dotted lines).}
    \label{fig:differentspinsspectralfunction}
\end{figure} 

Qualitative differences between off-diagonal matrix elements $B_{\alpha\beta}$ corresponding to energy eigenstates belonging to different spin sectors vs the same spin sector (considered in Sec.~\ref{sec:Off-diag-equalspins}) emerge when studying spectral functions. For $S_\alpha= S_\beta$, the results for $\omega=E_\alpha-E_\beta <0$ are identical to those for $\omega>0$ because $\hat B$ is Hermitian, which implies that $B_{\beta\alpha}=B^*_{\alpha\beta}$. For $S_\alpha \neq S_\beta$, the results for $\omega<0$ are independent from those for $\omega>0$. In that case, the Hermiticity of $\hat B$ implies that $f_B(\bar E,-\omega;\bar S,-\nu)=f^*_B(\bar E,\omega;\bar S,\nu)$. 

In Fig.~\ref{fig:differentspinsspectralfunction} we show results for the spectral function [see Eq.~\eqref{eq:spectralfunction}] vs $\omega$ (for both positive and negative frequencies) between energy eigenstates from spin sectors $(S_\alpha,S_\beta)$ with positive spin differences $\nu=S_\alpha-S_\beta=+2$, in the nonintegrable [Fig.~\ref{fig:differentspinsspectralfunction}(a)] and integrable [Fig.~\ref{fig:differentspinsspectralfunction}(b)] regimes. The spectral functions are not symmetric about $\omega=0$, and this is highlighted in the insets where we zoom into the region close to $\omega=0$. The inset in Fig.~\ref{fig:differentspinsspectralfunction}(a) further shows that in our finite systems the spectral function for $S_\alpha \neq S_\beta$ has its maximum for $\omega<0$, which contrasts the usual $\omega=0$ maximum found for $S_\alpha=S_\beta$. With increasing system size, we find that for $S_\alpha \neq S_\beta$ the maximum for $\omega<0$ moves toward $\omega=0$ [see the inset in Fig.~\ref{fig:differentspinsspectralfunction}(a)], so in the thermodynamic limit we expect it to occur at $\omega=0$. 

\begin{figure}[!t]
    \includegraphics[width=0.95\columnwidth]{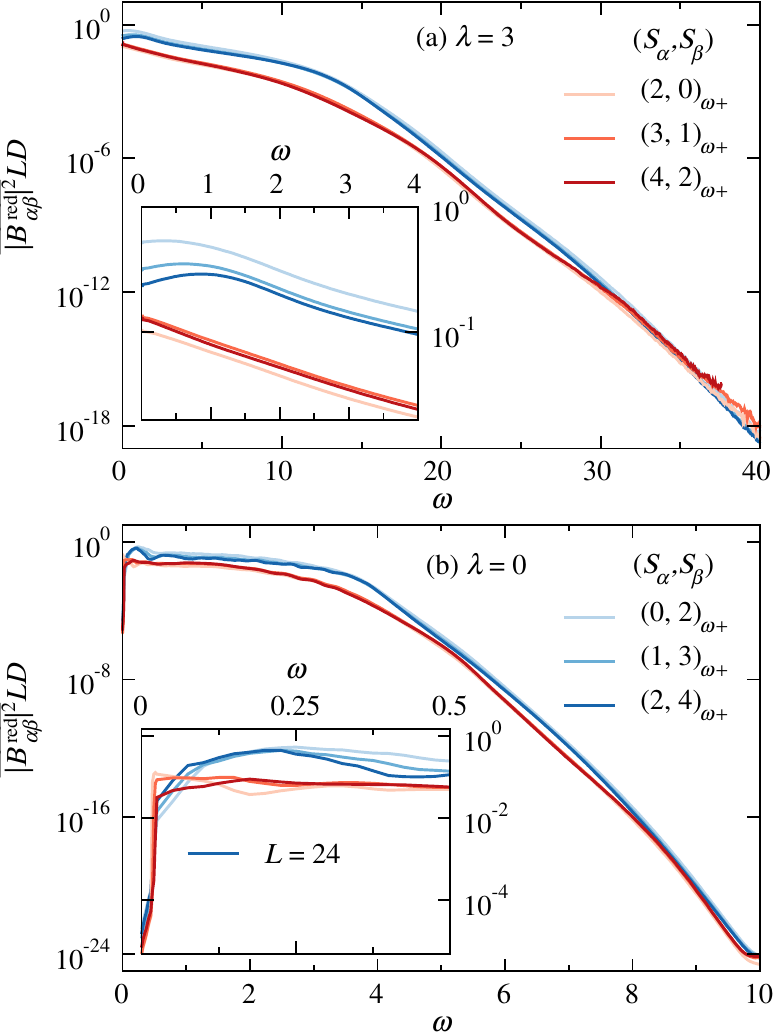}
    \vspace{-0.1cm}
    \caption{Same as Fig.~\ref{fig:differentspinsspectralfunction} but for the reduced matrix elements $B^\text{red}_{\alpha\beta}$ [see Eq.~\eqref{eq:MatrixelementRed}].}
    \label{fig:differentspins_reduced_spectralfunction}
\end{figure}       

In the context of the non-Abelian ETH, see Eq.~\eqref{eq:non-AbelianETH}, one is interested in the reduced matrix elements 
\begin{equation}\label{eq:MatrixelementRed}
    B^\text{red}_{\alpha\beta}\equiv \langle E_\alpha S_\alpha || B||E_\beta S_\beta\rangle\,.    
\end{equation}
They fully characterize the spectral functions for arbitrary values of $M_\alpha$ and $M_\beta$. For Hermitian spherical tensor operators of the form $\hat T^{(r)}_0$, positive and negative frequencies and positive and negative values of the spin differences are related via:
\begin{eqnarray}\label{eq:reducedrelation}
\langle E_\beta S_\beta ||T^{(r)}||E_\alpha S_\alpha\rangle &=& (-1)^{r}\frac{\sqrt{2S_\alpha+1}}{\sqrt{2S_\beta+1}} \\&& \times \langle E_\alpha S_\alpha || T^{(r)}||E_\beta S_\beta \rangle^*.\nonumber
\end{eqnarray}

In Fig.~\ref{fig:differentspins_reduced_spectralfunction}, we show the reduced spectral functions for $\omega\geq0$ corresponding to the results shown in Fig.~\ref{fig:differentspinsspectralfunction}. The $\omega=0$ jump between positive and negative spin differences (see insets in Fig.~\ref{fig:differentspins_reduced_spectralfunction}) is due to the prefactor in Eq.~\eqref{eq:reducedrelation}. Figure~\ref{fig:differentspins_reduced_spectralfunction} makes apparent that for $\omega\geq 0$ the spectral functions for positive and negative spin differences are different. This is the asymmetry we mentioned in the context of Fig.~\ref{fig:differentspinsspectralfunction} for a fixed spin difference when comparing positive and negative frequencies.

\begin{figure}
    \centering
    \includegraphics[width=0.98\columnwidth]{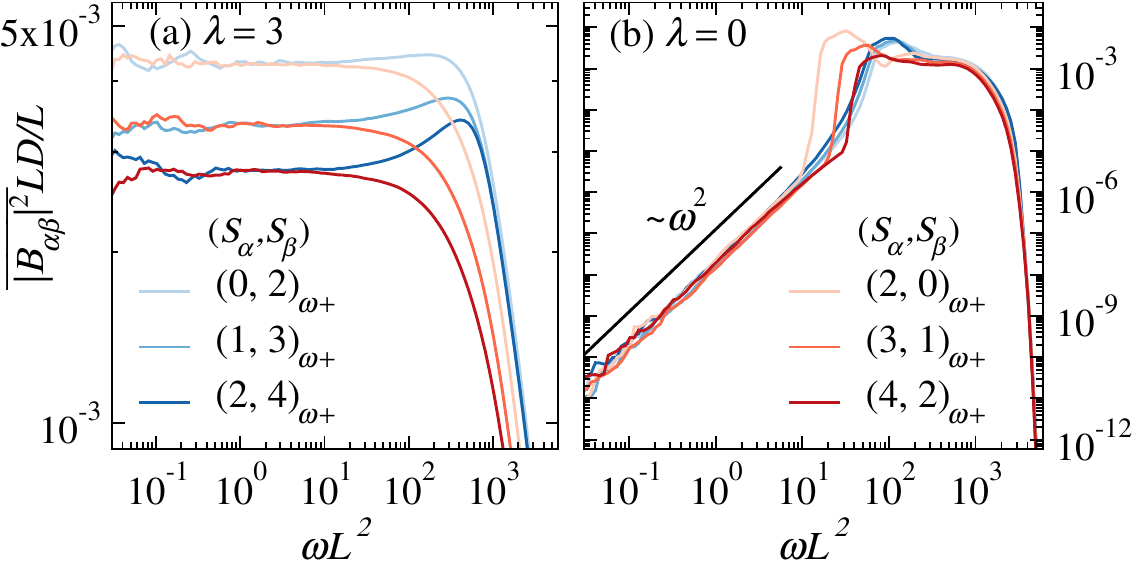}
    \vspace{-0.1cm}
    \caption{Low-frequency spectral function of the matrix elements $B_{\alpha\beta}=\langle E_\alpha S_\alpha M_\alpha|\hat B|E_\beta S_\beta M_\beta\rangle$ for: (a) the nonintegrable regime ($\lambda=3$), and (b) at integrability ($\lambda=0$) in chains with $L=24$. The solid black line in (b) is a guide to the eye showing $\omega^{2}$ scaling.}
    \label{fig:differentspinslowfrequencyspectralfunction}
\end{figure} 

We close our study of the matrix elements for $S_\alpha\neq S_\beta$ examining the spectral functions as $\omega\rightarrow 0$. As discussed for $S_\alpha= S_\beta$, one expects a plateau to develop in the nonintegrable regime, and the spectral function to vanish in the integrable regime for observables that preserve integrability. Our results in Fig.~\ref{fig:differentspinslowfrequencyspectralfunction} confirm those expectations. Remarkably, for a fixed $\bar S$ the plateaus in the nonintegrable regime [Fig.~\ref{fig:differentspinslowfrequencyspectralfunction}(a)] and the $\sim \omega^{2}$ vanishing spectral functions in the integrable regime [Fig.~\ref{fig:differentspinslowfrequencyspectralfunction}(b)] collapse onto each other for $\omega>0$ when $\nu'=-\nu$. This shows that as $\omega\rightarrow 0$ the spectral functions exhibit a symmetry that is associated with long-time transport that is not present at nonvanishing frequencies.           

\section{Summary}
\label{sec: Summary}

We studied the behavior of diagonal and off-diagonal matrix elements of spherical tensor observables in the eigenstates of the SU(2) symmetric extended spin-$\frac{1}{2}$ Heisenberg spin chain in the integrable ($\lambda=0$) and nonintegrable ($\lambda=3$) regimes. We considered observables that are SU(2) symmetric, which connect eigenstates within the same spin sector, and observables that break SU(2) symmetry, which connect eigenstates belonging to the same as well as different spin sectors.

For both types of observables, we examined the dependence of the diagonal matrix elements on the energy and spin densities. In the nonintegrable regime, focusing on sectors with vanishing and nonvanishing spin density $s=S/L$, we found evidence for the emergence of a smooth diagonal function in energy density consistent with the non-Abelian ETH, with fluctuations about the smooth function decaying exponentially with system size. Investigating the leading order energy dependence of the diagonal function (close to mean energy at infinite temperature $E_0$), we found contrasting behaviors when considering sectors with vanishing vs nonvanishing $s$. In the thermodynamic limit, the infinite temperature expectation value of the observables vanishes in the case of vanishing $s$, while it approaches an $O(1)$ number in the case of nonvanishing $s$. The coefficient of the linear in energy density term is independent of $s$ for the SU(2) symmetric observable, while it vanishes in the thermodynamic limit in the case of vanishing $s$ for the observable that breaks SU(2). Interestingly, when considering sectors with fixed spin and all possible magnetization values, the overlap of the observable with the Hamiltonian and its powers vanishes for all $s$ for the spherical tensor observables that break SU(2) ($\hat T^{(r)}_0$ with $r>0$), leading to a structureless (in the energy density) prediction for the diagonal function. We also probed the spin density dependence of the diagonal function at a fixed energy density, and found it to be consistent with the smooth dependence on $s$ as advanced by the non-Abelian ETH. 

Probing the behavior of off-diagonal matrix elements $O_{\alpha\beta}$ for eigenstates with the same spin $S_\alpha=S_\beta$ in the context of the non-Abelian ETH, we found results that are qualitatively similar to the ones obtained in the case of Abelian U(1) symmetry~\cite{leblond_2019,leblond_2020}. Notably, the spectral function $LD\overline{|O_{\alpha\beta}|^2}$ in the large-frequency limit vanishes exponentially with $\omega$ in the nonintegrable regime, while it exhibited a faster Gaussian decay $\propto e^{-a\omega^2}$ in the integrable regime. In the low-frequency limit, $LD\overline{|O_{\alpha\beta}|^2}$ in the nonintegrable regime plateaued to a value $\propto L$ for the observable that overlaps with the Hamiltonian, while the value scaled slower than $L$ for the observable whose overlap with the Hamiltonian vanishes in the thermodynamic limit for vanishing $s$. In the integrable regime, $LD\overline{|O_{\alpha\beta}|^2}$ was found to vanish as $\sim\omega^2$ consistent with the expectation for an integrability preserving observable. 

Qualitative differences between the behavior of off-diagonal elements $O_{\alpha\beta}$ connecting sectors with the same spin $S_\alpha= S_\beta$ vs different spins $S_\alpha\neq S_\beta$ were observed when studying the spectral functions. In the same spin case, the spectral functions $LD\overline{|O_{\alpha\beta}|^2}$ for the positive and negative frequencies coincide due to the Hermiticity of the observables. On the other hand, in the different spin case, the spectral functions for a fixed spin difference $\nu=S_\alpha-S_\beta$ were found to be asymmetric about $\omega=0$. In the low-frequency limit with $\omega>0$, the spectral functions at a fixed average spin $\bar S$ with spin differences $\nu$ and $-\nu$ collapse in the integrable and nonintegrable regimes, suggesting that transport at long times only depends on $|\nu|$.

It is remarkable that, despite the non-Abelian character of the SU(2) symmetry and the fact that $\hat{\vec S}^2$ is not an extensive quantity, all the eigenstate thermalization properties tested in this work behave as in systems with Abelian symmetries. All that SU(2) symmetry appears to do is introduce a further structure in the matrix elements as dictated by the Wigner–Eckart theorem. Our findings for eigenstate thermalization in the presence of SU(2) symmetry are consistent with the findings in Ref.~\cite{patil_2023} for the leading volume-law term in the average bipartite entanglement entropy of highly excited energy eigenstates of nonintegrable models with SU(2) symmetry at zero magnetization. Specifically, with the fact that the volume-law term is fully determined by the spin density and that it can be obtained assuming that the spin in the subsystem of interest and in its complement add to the full spin of the state (like for extensive quantities). The corrections to the latter approximation are subleading in the volume, and expected to be $O(1)$. An interesting open question is whether the lack of strict extensivity of $\hat{\vec S}^2$ can have an effect in the eigenstate thermalization phenomenon.

\acknowledgments
This work was supported by the National Science Foundation (NSF) Grant No.~PHY-2309146. The computations were done in the Institute for Computational and Data Sciences Roar supercomputer at Penn State. 

\appendix

\section{Non-spherical tensor observables}\label{app:offdiagETH}

In the main text, we explored the non-Abelian ETH in the context of spherical tensor operators $\hat A,\hat B$ of the form $\hat T^{(r)}_q$. As general operators $\hat O$ can be expanded in the basis of spherical tensor operators $\hat O= \sum_{r,q}C_{r,q}\hat T^{(r)}_q$, one expects ETH to also apply to $\hat O$ for the Hamiltonian in the nonintegrable regime. In this appendix, we examine the off-diagonal ETH for a non-spherical tensor operator, namely, $\hat C\equiv \frac{1}{L}\sum_{i=1}^L\hat S^z_{i}\hat S^z_{i+1}=-\sqrt{\frac{1}{3}}\hat T^{(0)}_0+\sqrt{\frac{2}{3}}\hat T^{(2)}_0$, where $\hat T^{(0)}_0=\hat A$ and $\hat T^{(2)}_0=\hat B$.

\begin{figure}
    \centering
    \includegraphics[width=0.96\columnwidth]{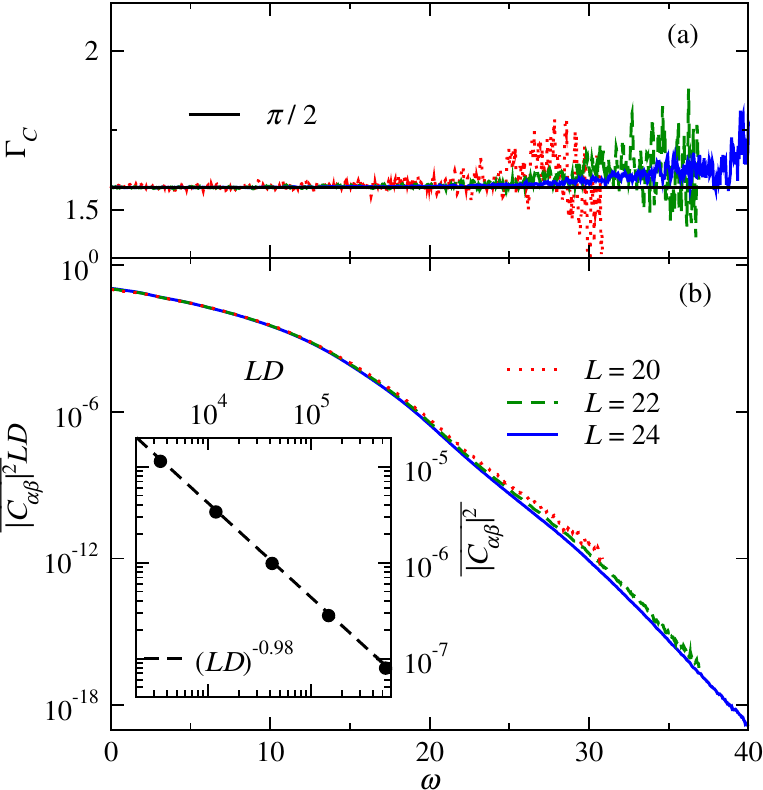}
    \vspace{-0.1cm}
    \caption{(a) $\Gamma_C$ [see Eq.~\eqref{eq:Gaussianitymeasure}] vs $\omega$ for spin $S=1$ in the nonintegrable regime ($\lambda=3$). The horizontal solid line marks $\Gamma=\pi/2$, corresponding to a Gaussian distribution. $\overline{|C_{\alpha\beta}|^2}$ and $\overline{|C_{\alpha\beta}|}^2$ in $\Gamma_C$ were computed using energy eigenstates with $|\bar E-E_0|/L\leqslant 0.025$, as a running average over $\omega$-windows of size $\Delta\omega=0.175$ for values of $\omega$ separated by $\delta\omega=0.025$. (b) Spectral functions vs $\omega$ for observable $\hat C$ and spin $S=1$ in the nonintegrable regime ($\lambda=3$). Inset in (b) shows the variance $\overline{|C_{\alpha\beta}|^2}$ vs $LD$ (averaged over $\omega\le10$). The black dashed line is a power law fit $\propto (LD)^\gamma$, with the fitting parameter $\gamma$ found to be close to $-1$.}
    \label{fig: Appendix-OffdiagETH}
\end{figure}

In Fig.~\ref{fig: Appendix-OffdiagETH}(a), we probe the (close to) Gaussian nature of the off-diagonal matrix elements of $\hat C$ in a sector with spin $S=1$ for the Hamiltonian in the nonintegrable regime ($\lambda=3$). Similar to the results for spherical tensor operators in Figs.~\ref{fig:Gaussianity}(a)--\ref{fig:Gaussianity}(d), $\Gamma_C\approx \pi/2$ (for $\omega\lesssim20$) and it remains close to $\pi/2$ for larger values of $\omega$ with increasing $L$. The inset in Fig.~\ref{fig: Appendix-OffdiagETH}(b) shows that the variance $\overline{|C_{\alpha\beta}|^2}$ decays as $(LD)^\gamma$ (with $\gamma\approx-1$), consistent with the expected $(LD)^{-1}$ scaling. Therefore, one can meaningfully compute the spectral function $\overline{|C_{\alpha\beta}|^2}LD$ vs $\omega$ as shown in Fig.~\ref{fig: Appendix-OffdiagETH}(b). The results for chains with $L=20,\,22,\,24$ collapse for $\omega\lesssim20$, and the gradual loss of collapse for higher $\omega$ (due to finite size effects) can be seen to occur at frequencies for which there is loss of Gaussianity (deviation of $\Gamma_C$ away from $\pi/2$) in Fig.~\ref{fig: Appendix-OffdiagETH}(a).

\section{Clebsch-Gordan coefficients}\label{app:CGcoeffs}

We study matrix elements of observables for $M_\alpha=M_\beta=q=0$ and $r=0,\,2$. The corresponding Clebsch-Gordan (CG) coefficients $\langle S_\alpha M_\alpha|S_\beta M_\beta; r q\rangle$ have the following form depending on the choice of $S_\alpha$ and $S_\beta$.
\begin{enumerate}
    \item For $S_\alpha=S_\beta=S$, the CG coefficient is $\langle S_\alpha M_\alpha|S_\beta M_\beta; r q\rangle=\langle S \, 0|S\, 0; r\, 0\rangle$.
    \begin{enumerate}
        \item For $S=O(1)$, since $r=O(1)$ the CG coefficient is $O(1)$.
        \item For $S=O(L)$, since $r=O(1)$ one can show that if $S=sL$ (with fixed spin density $s$) then 
        \begin{equation}
            \langle S\, 0|S\, 0; r\, 0\rangle =\begin{cases}
              1           & \,,r=0 \\  
            \frac{\sqrt{\pi}}{(-\frac{1}{2}-\frac{r}{2})!\frac{r}{2}!} +O(\frac{1}{L^2})& \,,r\ne 0 \text{ is even}\\
             0                       & \,,r \text{ is odd}
            \end{cases}
        \end{equation}
    \end{enumerate} 
    \item For $S_\alpha \neq S_\beta$ we only consider the case $S_\alpha=O(1)$ and $S_\beta=O(1)$. Since $r=O(1)$, the CG coefficient $\langle S_\alpha \, 0|S_\beta\, 0; r\, 0\rangle$ is $O(1)$. 
\end{enumerate}

\section{Infinite temperature expectation values}\label{app:Infinite-temperature-expectation-values}

In this appendix, we compute the infinite-temperature expectation values of the Hamiltonian $\hat H$ [Eq.~\eqref{eq:HamiltonianSpin1/2}] and the two observables $\hat A$ [Eq.~\eqref{eq:observableA}] and $\hat B$ [Eq.~\eqref{eq:observableB}], their joint moments, and the corresponding linear coefficients within sectors with a fixed total spin $S$ and zero total magnetization $M=0$. We focus on even $L$ chains with $L\ge6$. All the infinite temperature expectation values $\braket{\ldots}=\mathrm{Tr}(\ldots)/D_S$ are computed as traces over the total spin basis $\{\ket{S,M=0}_a\}$ with $a=1,\,\ldots,\,D_S$ labeling the $D_S$ eigenstates with total spin $S$ and magnetization $M=0$.

We start by noting that the infinite temperature expectation value $\epsilon_2=\braket{\hat{\vec{S}}_i \cdot \hat{\vec{S}}_{j}}$ for the product of two single-site spin operators $\hat{\vec{S}}_i$ and $\hat{\vec{S}}_j$,
\begin{equation}
    \epsilon_2=\frac{\mathrm{Tr}(\hat{\vec{S}}_i \cdot \hat{\vec{S}}_{j})}{D_S}=\sum_{a=1}^{D_S} \frac{\bra{S,0}_a \hat{\vec{S}}_i \cdot \hat{\vec{S}}_{j} \ket{S,0}_a}{D_S}\,,
\end{equation}
is independent of $i$ and $j\neq i$. In terms of $\epsilon_2$, the mean energy at infinite temperature can be written as, $E_0=\braket{\hat H}=-L(1+\lambda)\epsilon_2$. $\epsilon_2$ can be obtained calculating the infinite temperature expectation value of the total spin operator $\hat{\vec{S}}^2$:  
\begin{align}
    \braket{\hat{\vec{S}}^2}&=\sum_{i=1}^L \braket{\hat{\vec{S}}_i^2}+\sum_{i=1}^L\sum_{j\neq i}^L\braket{\hat{\vec{S}}_i\cdot \hat{\vec{S}}_j}\nonumber\\
    S(S+1)&=\frac{1}{2}\left(\frac{1}{2}+1\right)L+L(L-1)\epsilon_2\nonumber\\
    \implies \epsilon_2&=\frac{S(S+1)-\frac{3}{4}L}{L(L-1)}\,.
\end{align}
Therefore,
\begin{equation}\label{eq:AppC-<H>}
    E_0=\braket{\hat H}=(1+\lambda)\frac{\frac{3}{4}L-S(S+1)}{L-1}\,.
\end{equation}
Similarly, for observable $\hat A$ (which is proportional to $\hat H$ when $\lambda=0$),
\begin{equation}\label{eq:AppC-<A>}
    \braket{\hat A}=\frac{1}{\sqrt{3}}\frac{\frac{3}{4}L-S(S+1)}{L(L-1)}\,.
\end{equation}

On the other hand, for observable $\hat B$ we also need to compute the infinite temperature expectation value $\epsilon^z_2=\braket{\hat S_i^z \hat S_j^z}$ for the $z$-$z$ correlation term (which is, again, independent of $i$, $j\neq i$). $\epsilon^z_2$ can be obtained calculating the infinite temperature expectation value of the square of the total magnetization operator $\hat M=\sum_i \hat{S}^z_i$:
\begin{align}
    \braket{\hat M^2}&=\sum_{i=1}^L \braket{(\hat{S}^z_i)^2}+\sum_{i=1}^L\sum_{j\neq i}^L\braket{\hat S_i^z \hat S_j^z}\\
    0&=\frac{1}{4}L+L(L-1)\epsilon^z_2 \quad
    \implies \quad \epsilon^z_2=-\frac{1}{4(L-1)}\,.\nonumber
\end{align}
Therefore, the infinite temperature expectation value of the observable $\hat B=\sum_{i=1}^L (3\hat S_i^z \hat S_{i+1}^z-\hat{\vec{S}}_i \cdot \hat{\vec{S}}_{i+1})/\sqrt{6}L$ is:
\begin{equation}\label{eq:AppC-<B>}
    \braket{\hat B}=\frac{1}{\sqrt{6}}(3\epsilon^z_2-\epsilon_2)=-\frac{1}{\sqrt{6}}\frac{S(S+1)}{L(L-1)}\,.
\end{equation}

Next, we compute the joint moments $\braket{\hat O\hat H}$ of the observables $\hat O=\hat A,\hat B$ (and $\hat H$) with the Hamiltonian $\hat H$. To obtain the joint moments, we first need to compute the expectation values of two products of four single-site operators, namely, $\epsilon_4=\braket{(\hat{\vec{S}}_i \cdot \hat{\vec{S}}_{j})(\hat{\vec{S}}_{i'} \cdot \hat{\vec{S}}_{j'})}$ and $\epsilon^z_4=\braket{(\hat S_i^z \hat S_j^z)(\hat{\vec{S}}_{i'} \cdot \hat{\vec{S}}_{j'})}$ for $\{i',j'\}\neq\{i,j\}$. $\epsilon_4$ can be obtained calculating the infinite temperature expectation value of the square of the operator $\sum_i\sum_{j\neq i}\hat{\vec{S}}_i\cdot \hat{\vec{S}}_j$, 
\begin{eqnarray}\label{eq:AppC-(Si.Sj)^2}
    \left\langle\left(\sum_{i=1}^L\sum_{j\neq i}^L \hat{\vec{S}}_i\cdot \hat{\vec{S}}_j\right)^2\right\rangle&=&\left\langle\left(\hat{\vec{S}}^2-\sum_{i=1}^L \hat{\vec{S}}_i^2\right)^2\right\rangle\nonumber\\&=&\left[S(S+1)-\frac{3}{4}L\right]^2.
\end{eqnarray}
The left-hand side of Eq.~\eqref{eq:AppC-(Si.Sj)^2} contains a sum over three types of terms:
\begin{align}\label{eq:AppC-(Si.Sj)^2-kind1}
    2\sum_{i=1}^L\sum_{j\neq i}^L\left\langle\left(\hat{\vec{S}}_i\cdot \hat{\vec{S}}_j\right)^2\right\rangle&= 2\sum_{i=1}^L\sum_{j\neq i}^L \left\langle\frac{3}{16}-\frac{1}{2}\hat{\vec{S}}_i\cdot \hat{\vec{S}}_j\right\rangle\nonumber\\&= 2\frac{L!}{(L-2)!}\left(\frac{3}{16}-\frac{1}{2}\epsilon_2\right).
\end{align}
We also have terms of the form
\begin{align}\label{eq:AppC-(Si.Sj)^2-kind2}
    &4\sum_{i=1}^L\sum_{j\neq i}^L\sum_{j'\neq i,j}^L\braket{(\hat{\vec{S}}_i\cdot \hat{\vec{S}}_j)(\hat{\vec{S}}_i\cdot \hat{\vec{S}}_{j'})}\nonumber\\&= 4\sum_{i=1}^L\sum_{j\neq i}^L\sum_{j'\neq i,j}^L \frac{1}{4}\braket{\hat{\vec{S}}_j\cdot \hat{\vec{S}}_{j'}}+\frac{\mathrm{i}}{2}\braket{\hat S_i^z\hat S_j^x\hat S_{j'}^y+\ldots}\nonumber\\
    &\hspace{4.4cm}-\frac{\mathrm{i}}{2}\braket{\hat S_i^z\hat S_j^y\hat S_{j'}^x+\ldots}\nonumber\\
    &= 4\frac{L!}{(L-3)!}\left(\frac{1}{4}\epsilon_2\right),
\end{align}
where ``$\ldots$'' represents cyclic permutations. In the last step of Eq.~\eqref{eq:AppC-(Si.Sj)^2-kind2}, we used that the terms containing the product of three spin operators vanish under the $j\leftrightarrow j'$ exchange during the $j,\,j'$ summation. Lastly, we have terms
\begin{equation}\label{eq:AppC-(Si.Sj)^2-kind3}
    \sum_{i=1}^L\sum_{j\neq i}^L\sum_{i'\neq i,j}^L\sum_{j'\neq i,j,i'}^L\braket{(\hat{\vec{S}}_i \cdot \hat{\vec{S}}_{j})(\hat{\vec{S}}_{i'} \cdot \hat{\vec{S}}_{j'})}=\frac{L!}{(L-4)!}\epsilon_4\,.
\end{equation}
Plugging the results from Eqs.~\eqref{eq:AppC-(Si.Sj)^2-kind1},~\eqref{eq:AppC-(Si.Sj)^2-kind2},and~\eqref{eq:AppC-(Si.Sj)^2-kind3} into the left-hand side of Eq.~\eqref{eq:AppC-(Si.Sj)^2}, $\epsilon_4$ is found to be
\begin{equation}
    \epsilon_4=\frac{(\epsilon_2)^2L(L-1)-\epsilon_2(L-3)-3/8}{(L-2)(L-3)}\,.
\end{equation}
Similarly, $\epsilon^z_4$ can be obtained calculating the infinite temperature expectation value $\braket{(\hat M)^2(\sum_i\sum_{j\neq i}\hat{\vec{S}}_i\cdot \hat{\vec{S}}_j)}=0$:
\begin{align}\label{eq:AppC-M^2(Si.Sj)}
    &\left\langle\Big(\sum_{i=1}^L(\hat S_i^z)^2+\sum_{i=1}^L\sum_{j\neq i}^L \hat S_i^z \hat S_j^z\Big) \Big(\sum_{i'=1}^L\sum_{j'\neq i'}^L \hat{\vec{S}}_{i'}\cdot \hat{\vec{S}}_{j'}\Big)\right\rangle=0\nonumber\\&\left\langle\Big(\sum_{i=1}^L\sum_{j\neq i}^L \hat S_i^z \hat S_j^z\Big) \Big(\sum_{i'=1}^L\sum_{j'\neq i'}^L \hat{\vec{S}}_{i'}\cdot \hat{\vec{S}}_{j'}\Big)\right\rangle=-L^2(L-1)\frac{\epsilon_2}{4}\,,
\end{align}
where, the left-hand side of the last line of Eq.~\eqref{eq:AppC-M^2(Si.Sj)} has three types of terms that parallel those in Eq.~\eqref{eq:AppC-(Si.Sj)^2}:
\begin{align}\label{eq:AppC-M^2(Si.Sj)-kind1}
    &2\sum_{i=1}^L\sum_{j\neq i}^L\braket{(\hat S_i^z \hat S_j^z)(\hat{\vec{S}}_i\cdot \hat{\vec{S}}_j)}\nonumber\\&= 2\sum_{i=1}^L\sum_{j\neq i}^L \left\langle\frac{1}{16}-\frac{1}{4}\left(\hat{\vec{S}}_i\cdot \hat{\vec{S}}_j-\hat S_i^z \hat S_j^z\right)\right\rangle\nonumber\\&= 2\frac{L!}{(L-2)!}\left[\frac{1}{16}-\frac{1}{4}(\epsilon_2-\epsilon^z_2)\right]\,.
\end{align}
We also have terms of the form
\begin{align}\label{eq:AppC-M^2(Si.Sj)-kind2}
    &4\sum_{i=1}^L\sum_{j\neq i}^L\sum_{j'\neq i,j}^L\braket{(\hat S_i^z \hat S_j^z)(\hat{\vec{S}}_i\cdot \hat{\vec{S}}_{j'})}\nonumber\\&= 4\sum_{i=1}^L\sum_{j\neq i}^L\sum_{j'\neq i,j}^L \frac{1}{4}\braket{\hat S_j^z \hat S_{j'}^z}+\frac{\mathrm{i}}{2}\braket{\hat S_i^y\hat S_j^z\hat S_{j'}^x}-\frac{\mathrm{i}}{2}\braket{\hat S_i^x\hat S_j^z\hat S_{j'}^y}\nonumber\\
    &= 4\frac{L!}{(L-3)!}\left(\frac{1}{4}\epsilon^z_2\right),
\end{align}
where, in the last step of Eq.~\eqref{eq:AppC-M^2(Si.Sj)-kind2}, we used that the terms containing product of three spin operators vanish under the $i\leftrightarrow j'$ exchange during the $i,\,j'$ summation. Lastly, we have terms
\begin{equation}\label{eq:AppC-M^2(Si.Sj)-kind3}
    \sum_{i=1}^L\sum_{j\neq i}^L\sum_{i'\neq i,j}^L\sum_{j'\neq i,j,i'}^L\braket{(\hat S_i^z \hat S_j^z)(\hat{\vec{S}}_{i'} \cdot \hat{\vec{S}}_{j'})}=\frac{L!}{(L-4)!}\epsilon^z_4\,.
\end{equation}
Plugging the results from Eqs.~\eqref{eq:AppC-M^2(Si.Sj)-kind1},~\eqref{eq:AppC-M^2(Si.Sj)-kind2} and~\eqref{eq:AppC-M^2(Si.Sj)-kind3} into the left-hand side of Eq.~\eqref{eq:AppC-M^2(Si.Sj)}, $\epsilon^z_4$ is found to be
\begin{equation}
    \epsilon^z_4=-\frac{\epsilon_2(L-2)/4+\epsilon^z_2(L-3/2)+1/8}{(L-2)(L-3)}\,.
\end{equation}
Using the results for $\epsilon_4$ and $\epsilon^z_4$, in the following we evaluate the infinite temperature expectation values $\braket{\hat A \hat H}$,$\braket{\hat B\hat H}$ and $\braket{\hat H^2}$. For observable $\hat A$:
\begin{align}\label{eq:AppC-<AH>}
    \braket{\hat A \hat H}=&\left\langle\frac{\sum_{i=1}^L \hat{\vec{S}}_i \cdot \hat{\vec{S}}_{i+1}}{\sqrt{3}L}\left(\sum_{j=1}^L \hat{\vec{S}}_j \cdot \hat{\vec{S}}_{j+1}\right)\right\rangle \nonumber\\ &+ \left\langle\frac{\sum_{i=1}^L \hat{\vec{S}}_i \cdot \hat{\vec{S}}_{i+1}}{\sqrt{3}L}\left(\lambda\sum_{j=1}^L  \hat{\vec{S}}_j \cdot \hat{\vec{S}}_{j+2}\right)\right\rangle\nonumber\\
    =&\frac{1}{\sqrt{3}}\left[\left(\frac{3}{16}-\frac{\epsilon_2}{2}\right)+2\frac{\epsilon_2}{4}+(L-3)\epsilon_4\right]\nonumber\\&+\frac{\lambda}{\sqrt{3}}\left[4\frac{\epsilon_2}{4}+(L-4)\epsilon_4\right]\,,
\end{align}
where the two terms (with square brackets) in the result correspond to the product of $\hat A$ with the nearest and next-nearest terms in $\hat H$, respectively. Similarly,
\begin{align}\label{eq:AppC-<HH>}
    \braket{\hat H^2}=&\left\langle\left(\sum_{j=1}^L \hat{\vec{S}}_j \cdot \hat{\vec{S}}_{j+1} + \lambda \sum_{j=1}^L  \hat{\vec{S}}_j \cdot \hat{\vec{S}}_{j+2}\right)^2\right\rangle\nonumber\\
    =&(1+\lambda^2)L\left[\left(\frac{3}{16}-\frac{\epsilon_2}{2}\right)+2\frac{\epsilon_2}{4}+(L-3)\epsilon_4\right]\nonumber\\&+2\lambda L\left[4\frac{\epsilon_2}{4}+(L-4)\epsilon_4\right]\,.
\end{align}
For observable $\hat B$:
\begin{align}\label{eq:AppC-<BH>}
    \braket{\hat B \hat H}&=\left\langle\frac{\sum_{i=1}^L 3\hat S_i^z \hat S_{i+1}^z}{\sqrt{6}L}\hat H\right\rangle-\left\langle\frac{\sum_{i=1}^L \hat{\vec{S}}_i \cdot \hat{\vec{S}}_{i+1}}{\sqrt{6}L}\hat H\right\rangle\nonumber\\
    &=-\frac{3}{\sqrt{6}}\left[\left(\frac{1}{16}-\frac{1}{4}(\epsilon_2-\epsilon^z_2)\right)+2\frac{\epsilon^z_2}{4}+(L-3)\epsilon^z_4\right]\nonumber\\&\quad-\frac{3\lambda}{\sqrt{6}}\left[4\frac{\epsilon^z_2}{4}+(L-4)\epsilon^z_4\right]+\frac{1}{\sqrt{2}}\braket{\hat A\hat H}\,,
\end{align}
where the first two terms in the result (with square brackets) correspond to the product of the first term in $\hat B$ with $\hat H$, and the last term in the result $\braket{\hat A\hat H}/\sqrt{2}$ arises from the product of the second term in $\hat B$ with $\hat H$.

The coefficient of the linear (leading order nonvanishing) in energy density term in the expansion of the diagonal function $O(E_\alpha,S_\alpha)$ [Eq.~\eqref{eq:Taylor-expansion}] about $E_\alpha\approx E_0$ within a spin sector with $S_\alpha=S$ (and magnetization $M_\alpha=0$) is given by Eq.~\eqref{eq:linear-coeff}, which we rewrite below in terms of the moments
\begin{equation}\label{eq:AppC-linear-coeff}
    \frac{\partial O(E_\alpha,S)}{\partial (E_\alpha/L)}|_{E_\alpha=E_0}=L\frac{\braket{\hat O \hat H}_c}{\braket{\hat H \hat H}_c}=L\frac{\braket{\hat O \hat H}-\braket{\hat O} \braket{\hat H}}{\braket{\hat H^2}-\braket{\hat H}^2}\,.
\end{equation}
Using the results for the moments [Eqs.~\eqref{eq:AppC-<H>},~\eqref{eq:AppC-<A>} and~\eqref{eq:AppC-<B>}] and joint moments [Eqs.~\eqref{eq:AppC-<AH>},~\eqref{eq:AppC-<HH>} and~\eqref{eq:AppC-<BH>}] to evaluate the linear coefficient in Eq.~\eqref{eq:AppC-linear-coeff} for the two observables $\hat O=\hat A,\hat B$:
\begin{align}
    \frac{\partial A(E_\alpha,S)}{\partial (E_\alpha/L)}|_{E_\alpha=E_0}&=\frac{L-3-2\lambda}{\sqrt{3}[(L-3)(1+\lambda^2)-4\lambda]}\,, \\
    \frac{\partial B(E_\alpha,S)}{\partial (E_\alpha/L)}|_{E_\alpha=E_0}&=\frac{2\sqrt{2}S(S+1)}{4S(S+1)-3(L-2)^2}\\&\qquad\quad\times\frac{\partial A(E_\alpha,S)}{\partial (E_\alpha/L)}|_{E_\alpha=E_0}\nonumber.
\end{align}

\begin{thebibliography}{51}%
\makeatletter
\providecommand \@ifxundefined [1]{%
 \@ifx{#1\undefined}
}%
\providecommand \@ifnum [1]{%
 \ifnum #1\expandafter \@firstoftwo
 \else \expandafter \@secondoftwo
 \fi
}%
\providecommand \@ifx [1]{%
 \ifx #1\expandafter \@firstoftwo
 \else \expandafter \@secondoftwo
 \fi
}%
\providecommand \natexlab [1]{#1}%
\providecommand \enquote  [1]{``#1''}%
\providecommand \bibnamefont  [1]{#1}%
\providecommand \bibfnamefont [1]{#1}%
\providecommand \citenamefont [1]{#1}%
\providecommand \href@noop [0]{\@secondoftwo}%
\providecommand \href [0]{\begingroup \@sanitize@url \@href}%
\providecommand \@href[1]{\@@startlink{#1}\@@href}%
\providecommand \@@href[1]{\endgroup#1\@@endlink}%
\providecommand \@sanitize@url [0]{\catcode `\\12\catcode `\$12\catcode `\&12\catcode `\#12\catcode `\^12\catcode `\_12\catcode `\%12\relax}%
\providecommand \@@startlink[1]{}%
\providecommand \@@endlink[0]{}%
\providecommand \url  [0]{\begingroup\@sanitize@url \@url }%
\providecommand \@url [1]{\endgroup\@href {#1}{\urlprefix }}%
\providecommand \urlprefix  [0]{URL }%
\providecommand \Eprint [0]{\href }%
\providecommand \doibase [0]{https://doi.org/}%
\providecommand \selectlanguage [0]{\@gobble}%
\providecommand \bibinfo  [0]{\@secondoftwo}%
\providecommand \bibfield  [0]{\@secondoftwo}%
\providecommand \translation [1]{[#1]}%
\providecommand \BibitemOpen [0]{}%
\providecommand \bibitemStop [0]{}%
\providecommand \bibitemNoStop [0]{.\EOS\space}%
\providecommand \EOS [0]{\spacefactor3000\relax}%
\providecommand \BibitemShut  [1]{\csname bibitem#1\endcsname}%
\let\auto@bib@innerbib\@empty
\bibitem [{\citenamefont {D'Alessio}\ \emph {et~al.}(2016)\citenamefont {D'Alessio}, \citenamefont {Kafri}, \citenamefont {Polkovnikov},\ and\ \citenamefont {Rigol}}]{dalessio_quantum_2016}%
  \BibitemOpen
  \bibfield  {author} {\bibinfo {author} {\bibfnamefont {L.}~\bibnamefont {D'Alessio}}, \bibinfo {author} {\bibfnamefont {Y.}~\bibnamefont {Kafri}}, \bibinfo {author} {\bibfnamefont {A.}~\bibnamefont {Polkovnikov}},\ and\ \bibinfo {author} {\bibfnamefont {M.}~\bibnamefont {Rigol}},\ }\bibfield  {title} {\bibinfo {title} {From quantum chaos and eigenstate thermalization to statistical mechanics and thermodynamics},\ }\href {https://doi.org/10.1080/00018732.2016.1198134} {\bibfield  {journal} {\bibinfo  {journal} {Adv. Phys.}\ }\textbf {\bibinfo {volume} {65}},\ \bibinfo {pages} {239} (\bibinfo {year} {2016})}\BibitemShut {NoStop}%
\bibitem [{\citenamefont {Deutsch}(2018)}]{deutsch_18}%
  \BibitemOpen
  \bibfield  {author} {\bibinfo {author} {\bibfnamefont {J.~M.}\ \bibnamefont {Deutsch}},\ }\bibfield  {title} {\bibinfo {title} {Eigenstate thermalization hypothesis},\ }\href {https://doi.org/10.1088/1361-6633/aac9f1} {\bibfield  {journal} {\bibinfo  {journal} {Rep. Prog. Phys.}\ }\textbf {\bibinfo {volume} {81}},\ \bibinfo {pages} {082001} (\bibinfo {year} {2018})}\BibitemShut {NoStop}%
\bibitem [{\citenamefont {Mori}\ \emph {et~al.}(2018)\citenamefont {Mori}, \citenamefont {Ikeda}, \citenamefont {Kaminishi},\ and\ \citenamefont {Ueda}}]{mori_ikeda_18}%
  \BibitemOpen
  \bibfield  {author} {\bibinfo {author} {\bibfnamefont {T.}~\bibnamefont {Mori}}, \bibinfo {author} {\bibfnamefont {T.~N.}\ \bibnamefont {Ikeda}}, \bibinfo {author} {\bibfnamefont {E.}~\bibnamefont {Kaminishi}},\ and\ \bibinfo {author} {\bibfnamefont {M.}~\bibnamefont {Ueda}},\ }\bibfield  {title} {\bibinfo {title} {Thermalization and prethermalization in isolated quantum systems: {A} theoretical overview},\ }\href {https://doi.org/10.1088/1361-6455/aabcdf} {\bibfield  {journal} {\bibinfo  {journal} {J. Phys. B}\ }\textbf {\bibinfo {volume} {51}},\ \bibinfo {pages} {112001} (\bibinfo {year} {2018})}\BibitemShut {NoStop}%
\bibitem [{\citenamefont {Deutsch}(1991)}]{deutsch_1991}%
  \BibitemOpen
  \bibfield  {author} {\bibinfo {author} {\bibfnamefont {J.~M.}\ \bibnamefont {Deutsch}},\ }\bibfield  {title} {\bibinfo {title} {Quantum statistical mechanics in a closed system},\ }\href {https://doi.org/10.1103/PhysRevA.43.2046} {\bibfield  {journal} {\bibinfo  {journal} {Phys. Rev. A}\ }\textbf {\bibinfo {volume} {43}},\ \bibinfo {pages} {2046} (\bibinfo {year} {1991})}\BibitemShut {NoStop}%
\bibitem [{\citenamefont {Srednicki}(1994)}]{srednicki_1994}%
  \BibitemOpen
  \bibfield  {author} {\bibinfo {author} {\bibfnamefont {M.}~\bibnamefont {Srednicki}},\ }\bibfield  {title} {\bibinfo {title} {Chaos and quantum thermalization},\ }\href {https://doi.org/10.1103/PhysRevE.50.888} {\bibfield  {journal} {\bibinfo  {journal} {Phys. Rev. E}\ }\textbf {\bibinfo {volume} {50}},\ \bibinfo {pages} {888} (\bibinfo {year} {1994})}\BibitemShut {NoStop}%
\bibitem [{\citenamefont {Rigol}\ \emph {et~al.}(2008)\citenamefont {Rigol}, \citenamefont {Dunjko},\ and\ \citenamefont {Olshanii}}]{rigol_2008}%
  \BibitemOpen
  \bibfield  {author} {\bibinfo {author} {\bibfnamefont {M.}~\bibnamefont {Rigol}}, \bibinfo {author} {\bibfnamefont {V.}~\bibnamefont {Dunjko}},\ and\ \bibinfo {author} {\bibfnamefont {M.}~\bibnamefont {Olshanii}},\ }\bibfield  {title} {\bibinfo {title} {Thermalization and its mechanism for generic isolated quantum systems},\ }\href {https://doi.org/10.1038/nature06838} {\bibfield  {journal} {\bibinfo  {journal} {Nature}\ }\textbf {\bibinfo {volume} {452}},\ \bibinfo {pages} {854} (\bibinfo {year} {2008})}\BibitemShut {NoStop}%
\bibitem [{\citenamefont {Rigol}(2009{\natexlab{a}})}]{rigol_09a}%
  \BibitemOpen
  \bibfield  {author} {\bibinfo {author} {\bibfnamefont {M.}~\bibnamefont {Rigol}},\ }\bibfield  {title} {\bibinfo {title} {Breakdown of thermalization in finite one-dimensional systems},\ }\href {https://doi.org/10.1103/PhysRevLett.103.100403} {\bibfield  {journal} {\bibinfo  {journal} {Phys. Rev. Lett.}\ }\textbf {\bibinfo {volume} {103}},\ \bibinfo {pages} {100403} (\bibinfo {year} {2009}{\natexlab{a}})}\BibitemShut {NoStop}%
\bibitem [{\citenamefont {Rigol}(2009{\natexlab{b}})}]{rigol_09b}%
  \BibitemOpen
  \bibfield  {author} {\bibinfo {author} {\bibfnamefont {M.}~\bibnamefont {Rigol}},\ }\bibfield  {title} {\bibinfo {title} {Quantum quenches and thermalization in one-dimensional fermionic systems},\ }\href {https://doi.org/10.1103/PhysRevA.80.053607} {\bibfield  {journal} {\bibinfo  {journal} {Phys. Rev. A}\ }\textbf {\bibinfo {volume} {80}},\ \bibinfo {pages} {053607} (\bibinfo {year} {2009}{\natexlab{b}})}\BibitemShut {NoStop}%
\bibitem [{\citenamefont {Kim}\ \emph {et~al.}(2014)\citenamefont {Kim}, \citenamefont {Ikeda},\ and\ \citenamefont {Huse}}]{kim_testing_ETH_2014}%
  \BibitemOpen
  \bibfield  {author} {\bibinfo {author} {\bibfnamefont {H.}~\bibnamefont {Kim}}, \bibinfo {author} {\bibfnamefont {T.~N.}\ \bibnamefont {Ikeda}},\ and\ \bibinfo {author} {\bibfnamefont {D.~A.}\ \bibnamefont {Huse}},\ }\bibfield  {title} {\bibinfo {title} {Testing whether all eigenstates obey the eigenstate thermalization hypothesis},\ }\href {https://doi.org/10.1103/PhysRevE.90.052105} {\bibfield  {journal} {\bibinfo  {journal} {Phys. Rev. E}\ }\textbf {\bibinfo {volume} {90}},\ \bibinfo {pages} {052105} (\bibinfo {year} {2014})}\BibitemShut {NoStop}%
\bibitem [{\citenamefont {Beugeling}\ \emph {et~al.}(2014)\citenamefont {Beugeling}, \citenamefont {Moessner},\ and\ \citenamefont {Haque}}]{beugeling_2014}%
  \BibitemOpen
  \bibfield  {author} {\bibinfo {author} {\bibfnamefont {W.}~\bibnamefont {Beugeling}}, \bibinfo {author} {\bibfnamefont {R.}~\bibnamefont {Moessner}},\ and\ \bibinfo {author} {\bibfnamefont {M.}~\bibnamefont {Haque}},\ }\bibfield  {title} {\bibinfo {title} {Finite-size scaling of eigenstate thermalization},\ }\href {https://doi.org/10.1103/PhysRevE.89.042112} {\bibfield  {journal} {\bibinfo  {journal} {Phys. Rev. E}\ }\textbf {\bibinfo {volume} {89}},\ \bibinfo {pages} {042112} (\bibinfo {year} {2014})}\BibitemShut {NoStop}%
\bibitem [{\citenamefont {Beugeling}\ \emph {et~al.}(2015)\citenamefont {Beugeling}, \citenamefont {Moessner},\ and\ \citenamefont {Haque}}]{beugeling_offdiag_2015}%
  \BibitemOpen
  \bibfield  {author} {\bibinfo {author} {\bibfnamefont {W.}~\bibnamefont {Beugeling}}, \bibinfo {author} {\bibfnamefont {R.}~\bibnamefont {Moessner}},\ and\ \bibinfo {author} {\bibfnamefont {M.}~\bibnamefont {Haque}},\ }\bibfield  {title} {\bibinfo {title} {Off-diagonal matrix elements of local operators in many-body quantum systems},\ }\href {https://doi.org/10.1103/PhysRevE.91.012144} {\bibfield  {journal} {\bibinfo  {journal} {Phys. Rev. E}\ }\textbf {\bibinfo {volume} {91}},\ \bibinfo {pages} {012144} (\bibinfo {year} {2015})}\BibitemShut {NoStop}%
\bibitem [{\citenamefont {Mondaini}\ \emph {et~al.}(2016)\citenamefont {Mondaini}, \citenamefont {Fratus}, \citenamefont {Srednicki},\ and\ \citenamefont {Rigol}}]{mondaini_2016}%
  \BibitemOpen
  \bibfield  {author} {\bibinfo {author} {\bibfnamefont {R.}~\bibnamefont {Mondaini}}, \bibinfo {author} {\bibfnamefont {K.~R.}\ \bibnamefont {Fratus}}, \bibinfo {author} {\bibfnamefont {M.}~\bibnamefont {Srednicki}},\ and\ \bibinfo {author} {\bibfnamefont {M.}~\bibnamefont {Rigol}},\ }\bibfield  {title} {\bibinfo {title} {Eigenstate thermalization in the two-dimensional transverse field {I}sing model},\ }\href {https://doi.org/10.1103/PhysRevE.93.032104} {\bibfield  {journal} {\bibinfo  {journal} {Phys. Rev. E}\ }\textbf {\bibinfo {volume} {93}},\ \bibinfo {pages} {032104} (\bibinfo {year} {2016})}\BibitemShut {NoStop}%
\bibitem [{\citenamefont {Mondaini}\ and\ \citenamefont {Rigol}(2017)}]{mondaini_2017}%
  \BibitemOpen
  \bibfield  {author} {\bibinfo {author} {\bibfnamefont {R.}~\bibnamefont {Mondaini}}\ and\ \bibinfo {author} {\bibfnamefont {M.}~\bibnamefont {Rigol}},\ }\bibfield  {title} {\bibinfo {title} {Eigenstate thermalization in the two-dimensional transverse field {I}sing model. {II. O}ff-diagonal matrix elements of observables},\ }\href {https://doi.org/10.1103/PhysRevE.96.012157} {\bibfield  {journal} {\bibinfo  {journal} {Phys. Rev. E}\ }\textbf {\bibinfo {volume} {96}},\ \bibinfo {pages} {012157} (\bibinfo {year} {2017})}\BibitemShut {NoStop}%
\bibitem [{\citenamefont {LeBlond}\ \emph {et~al.}(2019)\citenamefont {LeBlond}, \citenamefont {Mallayya}, \citenamefont {Vidmar},\ and\ \citenamefont {Rigol}}]{leblond_2019}%
  \BibitemOpen
  \bibfield  {author} {\bibinfo {author} {\bibfnamefont {T.}~\bibnamefont {LeBlond}}, \bibinfo {author} {\bibfnamefont {K.}~\bibnamefont {Mallayya}}, \bibinfo {author} {\bibfnamefont {L.}~\bibnamefont {Vidmar}},\ and\ \bibinfo {author} {\bibfnamefont {M.}~\bibnamefont {Rigol}},\ }\bibfield  {title} {\bibinfo {title} {Entanglement and matrix elements of observables in interacting integrable systems},\ }\href {https://doi.org/10.1103/PhysRevE.100.062134} {\bibfield  {journal} {\bibinfo  {journal} {Phys. Rev. E}\ }\textbf {\bibinfo {volume} {100}},\ \bibinfo {pages} {062134} (\bibinfo {year} {2019})}\BibitemShut {NoStop}%
\bibitem [{\citenamefont {Jansen}\ \emph {et~al.}(2019)\citenamefont {Jansen}, \citenamefont {Stolpp}, \citenamefont {Vidmar},\ and\ \citenamefont {Heidrich-Meisner}}]{jansen_2019}%
  \BibitemOpen
  \bibfield  {author} {\bibinfo {author} {\bibfnamefont {D.}~\bibnamefont {Jansen}}, \bibinfo {author} {\bibfnamefont {J.}~\bibnamefont {Stolpp}}, \bibinfo {author} {\bibfnamefont {L.}~\bibnamefont {Vidmar}},\ and\ \bibinfo {author} {\bibfnamefont {F.}~\bibnamefont {Heidrich-Meisner}},\ }\bibfield  {title} {\bibinfo {title} {Eigenstate thermalization and quantum chaos in the {H}olstein polaron model},\ }\href {https://doi.org/10.1103/PhysRevB.99.155130} {\bibfield  {journal} {\bibinfo  {journal} {Phys. Rev. B}\ }\textbf {\bibinfo {volume} {99}},\ \bibinfo {pages} {155130} (\bibinfo {year} {2019})}\BibitemShut {NoStop}%
\bibitem [{\citenamefont {LeBlond}\ and\ \citenamefont {Rigol}(2020)}]{leblond_2020}%
  \BibitemOpen
  \bibfield  {author} {\bibinfo {author} {\bibfnamefont {T.}~\bibnamefont {LeBlond}}\ and\ \bibinfo {author} {\bibfnamefont {M.}~\bibnamefont {Rigol}},\ }\bibfield  {title} {\bibinfo {title} {Eigenstate thermalization for observables that break {Hamiltonian} symmetries and its counterpart in interacting integrable systems},\ }\href {https://doi.org/10.1103/PhysRevE.102.062113} {\bibfield  {journal} {\bibinfo  {journal} {Phys. Rev. E}\ }\textbf {\bibinfo {volume} {102}},\ \bibinfo {pages} {062113} (\bibinfo {year} {2020})}\BibitemShut {NoStop}%
\bibitem [{\citenamefont {Sch\"onle}\ \emph {et~al.}(2021)\citenamefont {Sch\"onle}, \citenamefont {Jansen}, \citenamefont {Heidrich-Meisner},\ and\ \citenamefont {Vidmar}}]{schonle_autocorrelation_2021}%
  \BibitemOpen
  \bibfield  {author} {\bibinfo {author} {\bibfnamefont {C.}~\bibnamefont {Sch\"onle}}, \bibinfo {author} {\bibfnamefont {D.}~\bibnamefont {Jansen}}, \bibinfo {author} {\bibfnamefont {F.}~\bibnamefont {Heidrich-Meisner}},\ and\ \bibinfo {author} {\bibfnamefont {L.}~\bibnamefont {Vidmar}},\ }\bibfield  {title} {\bibinfo {title} {Eigenstate thermalization hypothesis through the lens of autocorrelation functions},\ }\href {https://doi.org/10.1103/PhysRevB.103.235137} {\bibfield  {journal} {\bibinfo  {journal} {Phys. Rev. B}\ }\textbf {\bibinfo {volume} {103}},\ \bibinfo {pages} {235137} (\bibinfo {year} {2021})}\BibitemShut {NoStop}%
\bibitem [{\citenamefont {Wang}\ \emph {et~al.}(2024)\citenamefont {Wang}, \citenamefont {Zhu}, \citenamefont {Cui}, \citenamefont {Arg\"uello-Luengo}, \citenamefont {Lewenstein}, \citenamefont {Zhang}, \citenamefont {Sierant},\ and\ \citenamefont {Ran}}]{wang_2024}%
  \BibitemOpen
  \bibfield  {author} {\bibinfo {author} {\bibfnamefont {D.-Z.}\ \bibnamefont {Wang}}, \bibinfo {author} {\bibfnamefont {H.}~\bibnamefont {Zhu}}, \bibinfo {author} {\bibfnamefont {J.}~\bibnamefont {Cui}}, \bibinfo {author} {\bibfnamefont {J.}~\bibnamefont {Arg\"uello-Luengo}}, \bibinfo {author} {\bibfnamefont {M.}~\bibnamefont {Lewenstein}}, \bibinfo {author} {\bibfnamefont {G.-F.}\ \bibnamefont {Zhang}}, \bibinfo {author} {\bibfnamefont {P.}~\bibnamefont {Sierant}},\ and\ \bibinfo {author} {\bibfnamefont {S.-J.}\ \bibnamefont {Ran}},\ }\bibfield  {title} {\bibinfo {title} {Eigenstate thermalization and its breakdown in quantum spin chains with inhomogeneous interactions},\ }\href {https://doi.org/10.1103/PhysRevB.109.045139} {\bibfield  {journal} {\bibinfo  {journal} {Phys. Rev. B}\ }\textbf {\bibinfo {volume} {109}},\ \bibinfo {pages} {045139} (\bibinfo {year} {2024})}\BibitemShut {NoStop}%
\bibitem [{\citenamefont {Santos}\ and\ \citenamefont {Rigol}(2010{\natexlab{a}})}]{santos_rigol_10a}%
  \BibitemOpen
  \bibfield  {author} {\bibinfo {author} {\bibfnamefont {L.~F.}\ \bibnamefont {Santos}}\ and\ \bibinfo {author} {\bibfnamefont {M.}~\bibnamefont {Rigol}},\ }\bibfield  {title} {\bibinfo {title} {Onset of quantum chaos in one-dimensional bosonic and fermionic systems and its relation to thermalization},\ }\href {https://doi.org/10.1103/PhysRevE.81.036206} {\bibfield  {journal} {\bibinfo  {journal} {Phys. Rev. E}\ }\textbf {\bibinfo {volume} {81}},\ \bibinfo {pages} {036206} (\bibinfo {year} {2010}{\natexlab{a}})}\BibitemShut {NoStop}%
\bibitem [{\citenamefont {Santos}\ and\ \citenamefont {Rigol}(2010{\natexlab{b}})}]{santos_rigol_10b}%
  \BibitemOpen
  \bibfield  {author} {\bibinfo {author} {\bibfnamefont {L.~F.}\ \bibnamefont {Santos}}\ and\ \bibinfo {author} {\bibfnamefont {M.}~\bibnamefont {Rigol}},\ }\bibfield  {title} {\bibinfo {title} {Localization and the effects of symmetries in the thermalization properties of one-dimensional quantum systems},\ }\href {https://doi.org/10.1103/PhysRevE.82.031130} {\bibfield  {journal} {\bibinfo  {journal} {Phys. Rev. E}\ }\textbf {\bibinfo {volume} {82}},\ \bibinfo {pages} {031130} (\bibinfo {year} {2010}{\natexlab{b}})}\BibitemShut {NoStop}%
\bibitem [{\citenamefont {Mondaini}\ \emph {et~al.}(2018)\citenamefont {Mondaini}, \citenamefont {Mallayya}, \citenamefont {Santos},\ and\ \citenamefont {Rigol}}]{mondaini_mallayya_18}%
  \BibitemOpen
  \bibfield  {author} {\bibinfo {author} {\bibfnamefont {R.}~\bibnamefont {Mondaini}}, \bibinfo {author} {\bibfnamefont {K.}~\bibnamefont {Mallayya}}, \bibinfo {author} {\bibfnamefont {L.~F.}\ \bibnamefont {Santos}},\ and\ \bibinfo {author} {\bibfnamefont {M.}~\bibnamefont {Rigol}},\ }\bibfield  {title} {\bibinfo {title} {Comment on ``{S}ystematic construction of counterexamples to the eigenstate thermalization hypothesis''},\ }\href {https://doi.org/10.1103/PhysRevLett.121.038901} {\bibfield  {journal} {\bibinfo  {journal} {Phys. Rev. Lett.}\ }\textbf {\bibinfo {volume} {121}},\ \bibinfo {pages} {038901} (\bibinfo {year} {2018})}\BibitemShut {NoStop}%
\bibitem [{\citenamefont {Halpern}\ \emph {et~al.}(2016)\citenamefont {Halpern}, \citenamefont {Faist}, \citenamefont {Oppenheim},\ and\ \citenamefont {Winter}}]{halpern_2016}%
  \BibitemOpen
  \bibfield  {author} {\bibinfo {author} {\bibfnamefont {N.~Y.}\ \bibnamefont {Halpern}}, \bibinfo {author} {\bibfnamefont {P.}~\bibnamefont {Faist}}, \bibinfo {author} {\bibfnamefont {J.}~\bibnamefont {Oppenheim}},\ and\ \bibinfo {author} {\bibfnamefont {A.}~\bibnamefont {Winter}},\ }\bibfield  {title} {\bibinfo {title} {Microcanonical and resource-theoretic derivations of the thermal state of a quantum system with noncommuting charges},\ }\href {https://doi.org/10.1038/ncomms12051} {\bibfield  {journal} {\bibinfo  {journal} {Nat. Commun.}\ }\textbf {\bibinfo {volume} {7}},\ \bibinfo {pages} {12051} (\bibinfo {year} {2016})}\BibitemShut {NoStop}%
\bibitem [{\citenamefont {Guryanova}\ \emph {et~al.}(2016)\citenamefont {Guryanova}, \citenamefont {Popescu}, \citenamefont {Short}, \citenamefont {Silva},\ and\ \citenamefont {Skrzypczyk}}]{guryanova_2016}%
  \BibitemOpen
  \bibfield  {author} {\bibinfo {author} {\bibfnamefont {Y.}~\bibnamefont {Guryanova}}, \bibinfo {author} {\bibfnamefont {S.}~\bibnamefont {Popescu}}, \bibinfo {author} {\bibfnamefont {A.~J.}\ \bibnamefont {Short}}, \bibinfo {author} {\bibfnamefont {R.}~\bibnamefont {Silva}},\ and\ \bibinfo {author} {\bibfnamefont {P.}~\bibnamefont {Skrzypczyk}},\ }\bibfield  {title} {\bibinfo {title} {Thermodynamics of quantum systems with multiple conserved quantities},\ }\href {https://doi.org/10.1038/ncomms12049} {\bibfield  {journal} {\bibinfo  {journal} {Nat. Commun.}\ }\textbf {\bibinfo {volume} {7}},\ \bibinfo {pages} {12049} (\bibinfo {year} {2016})}\BibitemShut {NoStop}%
\bibitem [{\citenamefont {Popescu}\ \emph {et~al.}(2020)\citenamefont {Popescu}, \citenamefont {Sainz}, \citenamefont {Short},\ and\ \citenamefont {Winter}}]{popescu_2020}%
  \BibitemOpen
  \bibfield  {author} {\bibinfo {author} {\bibfnamefont {S.}~\bibnamefont {Popescu}}, \bibinfo {author} {\bibfnamefont {A.~B.}\ \bibnamefont {Sainz}}, \bibinfo {author} {\bibfnamefont {A.~J.}\ \bibnamefont {Short}},\ and\ \bibinfo {author} {\bibfnamefont {A.}~\bibnamefont {Winter}},\ }\bibfield  {title} {\bibinfo {title} {Reference frames which separately store noncommuting conserved quantities},\ }\href {https://doi.org/10.1103/PhysRevLett.125.090601} {\bibfield  {journal} {\bibinfo  {journal} {Phys. Rev. Lett.}\ }\textbf {\bibinfo {volume} {125}},\ \bibinfo {pages} {090601} (\bibinfo {year} {2020})}\BibitemShut {NoStop}%
\bibitem [{\citenamefont {Yunger~Halpern}\ \emph {et~al.}(2020)\citenamefont {Yunger~Halpern}, \citenamefont {Beverland},\ and\ \citenamefont {Kalev}}]{halpern_2020}%
  \BibitemOpen
  \bibfield  {author} {\bibinfo {author} {\bibfnamefont {N.}~\bibnamefont {Yunger~Halpern}}, \bibinfo {author} {\bibfnamefont {M.~E.}\ \bibnamefont {Beverland}},\ and\ \bibinfo {author} {\bibfnamefont {A.}~\bibnamefont {Kalev}},\ }\bibfield  {title} {\bibinfo {title} {Noncommuting conserved charges in quantum many-body thermalization},\ }\href {https://doi.org/10.1103/PhysRevE.101.042117} {\bibfield  {journal} {\bibinfo  {journal} {Phys. Rev. E}\ }\textbf {\bibinfo {volume} {101}},\ \bibinfo {pages} {042117} (\bibinfo {year} {2020})}\BibitemShut {NoStop}%
\bibitem [{\citenamefont {Manzano}\ \emph {et~al.}(2022)\citenamefont {Manzano}, \citenamefont {Parrondo},\ and\ \citenamefont {Landi}}]{manzano_2022}%
  \BibitemOpen
  \bibfield  {author} {\bibinfo {author} {\bibfnamefont {G.}~\bibnamefont {Manzano}}, \bibinfo {author} {\bibfnamefont {J.~M.}\ \bibnamefont {Parrondo}},\ and\ \bibinfo {author} {\bibfnamefont {G.~T.}\ \bibnamefont {Landi}},\ }\bibfield  {title} {\bibinfo {title} {Non-abelian quantum transport and thermosqueezing effects},\ }\href {https://doi.org/10.1103/PRXQuantum.3.010304} {\bibfield  {journal} {\bibinfo  {journal} {PRX Quantum}\ }\textbf {\bibinfo {volume} {3}},\ \bibinfo {pages} {010304} (\bibinfo {year} {2022})}\BibitemShut {NoStop}%
\bibitem [{\citenamefont {Kranzl}\ \emph {et~al.}(2023)\citenamefont {Kranzl}, \citenamefont {Lasek}, \citenamefont {Joshi}, \citenamefont {Kalev}, \citenamefont {Blatt}, \citenamefont {Roos},\ and\ \citenamefont {Yunger~Halpern}}]{kranzl_2023}%
  \BibitemOpen
  \bibfield  {author} {\bibinfo {author} {\bibfnamefont {F.}~\bibnamefont {Kranzl}}, \bibinfo {author} {\bibfnamefont {A.}~\bibnamefont {Lasek}}, \bibinfo {author} {\bibfnamefont {M.~K.}\ \bibnamefont {Joshi}}, \bibinfo {author} {\bibfnamefont {A.}~\bibnamefont {Kalev}}, \bibinfo {author} {\bibfnamefont {R.}~\bibnamefont {Blatt}}, \bibinfo {author} {\bibfnamefont {C.~F.}\ \bibnamefont {Roos}},\ and\ \bibinfo {author} {\bibfnamefont {N.}~\bibnamefont {Yunger~Halpern}},\ }\bibfield  {title} {\bibinfo {title} {Experimental observation of thermalization with noncommuting charges},\ }\href {https://doi.org/10.1103/PRXQuantum.4.020318} {\bibfield  {journal} {\bibinfo  {journal} {PRX Quantum}\ }\textbf {\bibinfo {volume} {4}},\ \bibinfo {pages} {020318} (\bibinfo {year} {2023})}\BibitemShut {NoStop}%
\bibitem [{\citenamefont {Majidy}\ \emph {et~al.}(2023{\natexlab{a}})\citenamefont {Majidy}, \citenamefont {Braasch}, \citenamefont {Lasek}, \citenamefont {Upadhyaya}, \citenamefont {Kalev},\ and\ \citenamefont {Yunger~Halpern}}]{majidy_review_2023}%
  \BibitemOpen
  \bibfield  {author} {\bibinfo {author} {\bibfnamefont {S.}~\bibnamefont {Majidy}}, \bibinfo {author} {\bibfnamefont {W.~F.}\ \bibnamefont {Braasch}}, \bibinfo {author} {\bibfnamefont {A.}~\bibnamefont {Lasek}}, \bibinfo {author} {\bibfnamefont {T.}~\bibnamefont {Upadhyaya}}, \bibinfo {author} {\bibfnamefont {A.}~\bibnamefont {Kalev}},\ and\ \bibinfo {author} {\bibfnamefont {N.}~\bibnamefont {Yunger~Halpern}},\ }\bibfield  {title} {\bibinfo {title} {Noncommuting conserved charges in quantum thermodynamics and beyond},\ }\href {https://doi.org/10.1038/s42254-023-00641-9} {\bibfield  {journal} {\bibinfo  {journal} {Nature Reviews Physics}\ }\textbf {\bibinfo {volume} {5}},\ \bibinfo {pages} {689} (\bibinfo {year} {2023}{\natexlab{a}})}\BibitemShut {NoStop}%
\bibitem [{\citenamefont {Patil}\ \emph {et~al.}(2023)\citenamefont {Patil}, \citenamefont {Hackl}, \citenamefont {Fagan},\ and\ \citenamefont {Rigol}}]{patil_2023}%
  \BibitemOpen
  \bibfield  {author} {\bibinfo {author} {\bibfnamefont {R.}~\bibnamefont {Patil}}, \bibinfo {author} {\bibfnamefont {L.}~\bibnamefont {Hackl}}, \bibinfo {author} {\bibfnamefont {G.~R.}\ \bibnamefont {Fagan}},\ and\ \bibinfo {author} {\bibfnamefont {M.}~\bibnamefont {Rigol}},\ }\bibfield  {title} {\bibinfo {title} {Average pure-state entanglement entropy in spin systems with {SU(2)} symmetry},\ }\href {https://doi.org/10.1103/PhysRevB.108.245101} {\bibfield  {journal} {\bibinfo  {journal} {Phys. Rev. B}\ }\textbf {\bibinfo {volume} {108}},\ \bibinfo {pages} {245101} (\bibinfo {year} {2023})}\BibitemShut {NoStop}%
\bibitem [{\citenamefont {Majidy}\ \emph {et~al.}(2023{\natexlab{b}})\citenamefont {Majidy}, \citenamefont {Lasek}, \citenamefont {Huse},\ and\ \citenamefont {Yunger~Halpern}}]{majidy_entanglement_2023}%
  \BibitemOpen
  \bibfield  {author} {\bibinfo {author} {\bibfnamefont {S.}~\bibnamefont {Majidy}}, \bibinfo {author} {\bibfnamefont {A.}~\bibnamefont {Lasek}}, \bibinfo {author} {\bibfnamefont {D.~A.}\ \bibnamefont {Huse}},\ and\ \bibinfo {author} {\bibfnamefont {N.}~\bibnamefont {Yunger~Halpern}},\ }\bibfield  {title} {\bibinfo {title} {Non-abelian symmetry can increase entanglement entropy},\ }\href {https://doi.org/10.1103/PhysRevB.107.045102} {\bibfield  {journal} {\bibinfo  {journal} {Phys. Rev. B}\ }\textbf {\bibinfo {volume} {107}},\ \bibinfo {pages} {045102} (\bibinfo {year} {2023}{\natexlab{b}})}\BibitemShut {NoStop}%
\bibitem [{\citenamefont {Bianchi}\ \emph {et~al.}(2024)\citenamefont {Bianchi}, \citenamefont {Dona},\ and\ \citenamefont {Kumar}}]{bianchi_2024}%
  \BibitemOpen
  \bibfield  {author} {\bibinfo {author} {\bibfnamefont {E.}~\bibnamefont {Bianchi}}, \bibinfo {author} {\bibfnamefont {P.}~\bibnamefont {Dona}},\ and\ \bibinfo {author} {\bibfnamefont {R.}~\bibnamefont {Kumar}},\ }\href {https://arxiv.org/abs/2405.00597} {\bibinfo {title} {Non-abelian symmetry-resolved entanglement entropy}} (\bibinfo {year} {2024}),\ \Eprint {https://arxiv.org/abs/2405.00597} {arXiv:2405.00597 [quant-ph]} \BibitemShut {NoStop}%
\bibitem [{\citenamefont {Murthy}\ \emph {et~al.}(2023)\citenamefont {Murthy}, \citenamefont {Babakhani}, \citenamefont {Iniguez}, \citenamefont {Srednicki},\ and\ \citenamefont {Yunger~Halpern}}]{murthy_nonabelian_2023}%
  \BibitemOpen
  \bibfield  {author} {\bibinfo {author} {\bibfnamefont {C.}~\bibnamefont {Murthy}}, \bibinfo {author} {\bibfnamefont {A.}~\bibnamefont {Babakhani}}, \bibinfo {author} {\bibfnamefont {F.}~\bibnamefont {Iniguez}}, \bibinfo {author} {\bibfnamefont {M.}~\bibnamefont {Srednicki}},\ and\ \bibinfo {author} {\bibfnamefont {N.}~\bibnamefont {Yunger~Halpern}},\ }\bibfield  {title} {\bibinfo {title} {Non-abelian eigenstate thermalization hypothesis},\ }\href {https://doi.org/10.1103/PhysRevLett.130.140402} {\bibfield  {journal} {\bibinfo  {journal} {Phys. Rev. Lett.}\ }\textbf {\bibinfo {volume} {130}},\ \bibinfo {pages} {140402} (\bibinfo {year} {2023})}\BibitemShut {NoStop}%
\bibitem [{\citenamefont {Noh}(2023)}]{noh_2023}%
  \BibitemOpen
  \bibfield  {author} {\bibinfo {author} {\bibfnamefont {J.~D.}\ \bibnamefont {Noh}},\ }\bibfield  {title} {\bibinfo {title} {Eigenstate thermalization hypothesis in two-dimensional {$XXZ$} model with or without {SU(2)} symmetry},\ }\href {https://doi.org/10.1103/PhysRevE.107.014130} {\bibfield  {journal} {\bibinfo  {journal} {Phys. Rev. E}\ }\textbf {\bibinfo {volume} {107}},\ \bibinfo {pages} {014130} (\bibinfo {year} {2023})}\BibitemShut {NoStop}%
\bibitem [{\citenamefont {Lasek}\ \emph {et~al.}(2024)\citenamefont {Lasek}, \citenamefont {Noh}, \citenamefont {LeSchack},\ and\ \citenamefont {Halpern}}]{lasek_24}%
  \BibitemOpen
  \bibfield  {author} {\bibinfo {author} {\bibfnamefont {A.}~\bibnamefont {Lasek}}, \bibinfo {author} {\bibfnamefont {J.~D.}\ \bibnamefont {Noh}}, \bibinfo {author} {\bibfnamefont {J.}~\bibnamefont {LeSchack}},\ and\ \bibinfo {author} {\bibfnamefont {N.~Y.}\ \bibnamefont {Halpern}},\ }\href {https://arxiv.org/abs/2412.07838} {\bibinfo {title} {Numerical evidence for the non-abelian eigenstate thermalization hypothesis}} (\bibinfo {year} {2024}),\ \Eprint {https://arxiv.org/abs/2412.07838} {arXiv:2412.07838 [quant-ph]} \BibitemShut {NoStop}%
\bibitem [{\citenamefont {Shankar}(1994)}]{Shankar1994}%
  \BibitemOpen
  \bibfield  {author} {\bibinfo {author} {\bibfnamefont {R.}~\bibnamefont {Shankar}},\ }\href {https://doi.org/10.1007/978-1-4757-0576-8} {\emph {\bibinfo {title} {Principles of Quantum Mechanics}}}\ (\bibinfo  {publisher} {Springer US},\ \bibinfo {year} {1994})\BibitemShut {NoStop}%
\bibitem [{\citenamefont {Sakurai}\ and\ \citenamefont {Napolitano}(2017)}]{Sakurai2017}%
  \BibitemOpen
  \bibfield  {author} {\bibinfo {author} {\bibfnamefont {J.~J.}\ \bibnamefont {Sakurai}}\ and\ \bibinfo {author} {\bibfnamefont {J.}~\bibnamefont {Napolitano}},\ }\href {https://doi.org/10.1017/9781108499996} {\emph {\bibinfo {title} {Modern Quantum Mechanics}}}\ (\bibinfo  {publisher} {Cambridge University Press},\ \bibinfo {year} {2017})\BibitemShut {NoStop}%
\bibitem [{Note1()}]{Note1}%
  \BibitemOpen
  \bibinfo {note} {Since $\protect \hat B$ is a $\protect \hat T^{(2)}_0$ operator, its diagonal matrix elements vanish between states with $S=0$. Therefore, we fix $S=1$ so that the diagonal matrix elements are nonvanishing yet, in the limit $L = \infty $, the spin density $s = 0$ as for $S=0$ considered for observable $\protect \hat A$.}\BibitemShut {Stop}%
\bibitem [{\citenamefont {Capizzi}\ \emph {et~al.}(2024)\citenamefont {Capizzi}, \citenamefont {Wang}, \citenamefont {Xu}, \citenamefont {Mazza},\ and\ \citenamefont {Poletti}}]{capizzi_poletti_24}%
  \BibitemOpen
  \bibfield  {author} {\bibinfo {author} {\bibfnamefont {L.}~\bibnamefont {Capizzi}}, \bibinfo {author} {\bibfnamefont {J.}~\bibnamefont {Wang}}, \bibinfo {author} {\bibfnamefont {X.}~\bibnamefont {Xu}}, \bibinfo {author} {\bibfnamefont {L.}~\bibnamefont {Mazza}},\ and\ \bibinfo {author} {\bibfnamefont {D.}~\bibnamefont {Poletti}},\ }\href {https://arxiv.org/abs/2405.16975} {\bibinfo {title} {Hydrodynamics and the eigenstate thermalization hypothesis}} (\bibinfo {year} {2024}),\ \Eprint {https://arxiv.org/abs/2405.16975} {arXiv:2405.16975 [quant-ph]} \BibitemShut {NoStop}%
\bibitem [{\citenamefont {Varshalovich}\ \emph {et~al.}(1988)\citenamefont {Varshalovich}, \citenamefont {Moskalev},\ and\ \citenamefont {Khersonskii}}]{clebsch_gordan_formula_book_1988}%
  \BibitemOpen
  \bibfield  {author} {\bibinfo {author} {\bibfnamefont {D.~A.}\ \bibnamefont {Varshalovich}}, \bibinfo {author} {\bibfnamefont {A.~N.}\ \bibnamefont {Moskalev}},\ and\ \bibinfo {author} {\bibfnamefont {V.~K.}\ \bibnamefont {Khersonskii}},\ }\href {https://doi.org/10.1142/0270} {\emph {\bibinfo {title} {Quantum Theory of Angular Momentum}}}\ (\bibinfo  {publisher} {WORLD SCIENTIFIC},\ \bibinfo {year} {1988})\BibitemShut {NoStop}%
\bibitem [{\citenamefont {Mierzejewski}\ and\ \citenamefont {Vidmar}(2020)}]{mierzejewski_vidmar_20}%
  \BibitemOpen
  \bibfield  {author} {\bibinfo {author} {\bibfnamefont {M.}~\bibnamefont {Mierzejewski}}\ and\ \bibinfo {author} {\bibfnamefont {L.}~\bibnamefont {Vidmar}},\ }\bibfield  {title} {\bibinfo {title} {Quantitative impact of integrals of motion on the eigenstate thermalization hypothesis},\ }\href {https://doi.org/10.1103/PhysRevLett.124.040603} {\bibfield  {journal} {\bibinfo  {journal} {Phys. Rev. Lett.}\ }\textbf {\bibinfo {volume} {124}},\ \bibinfo {pages} {040603} (\bibinfo {year} {2020})}\BibitemShut {NoStop}%
\bibitem [{\citenamefont {\L{}yd\ifmmode~\dot{z}\else \.{z}\fi{}ba}\ \emph {et~al.}(2024)\citenamefont {\L{}yd\ifmmode~\dot{z}\else \.{z}\fi{}ba}, \citenamefont {\ifmmode \acute{S}\else \'{S}\fi{}wi\ifmmode~\mbox{\k{e}}\else \k{e}\fi{}tek}, \citenamefont {Mierzejewski}, \citenamefont {Rigol},\ and\ \citenamefont {Vidmar}}]{patrycja_rafal_24}%
  \BibitemOpen
  \bibfield  {author} {\bibinfo {author} {\bibfnamefont {P.}~\bibnamefont {\L{}yd\ifmmode~\dot{z}\else \.{z}\fi{}ba}}, \bibinfo {author} {\bibfnamefont {R.}~\bibnamefont {\ifmmode \acute{S}\else \'{S}\fi{}wi\ifmmode~\mbox{\k{e}}\else \k{e}\fi{}tek}}, \bibinfo {author} {\bibfnamefont {M.}~\bibnamefont {Mierzejewski}}, \bibinfo {author} {\bibfnamefont {M.}~\bibnamefont {Rigol}},\ and\ \bibinfo {author} {\bibfnamefont {L.}~\bibnamefont {Vidmar}},\ }\bibfield  {title} {\bibinfo {title} {Normal weak eigenstate thermalization},\ }\href {https://doi.org/10.1103/PhysRevB.110.104202} {\bibfield  {journal} {\bibinfo  {journal} {Phys. Rev. B}\ }\textbf {\bibinfo {volume} {110}},\ \bibinfo {pages} {104202} (\bibinfo {year} {2024})}\BibitemShut {NoStop}%
\bibitem [{Note2()}]{Note2}%
  \BibitemOpen
  \bibinfo {note} {The fact that correlations must exist between matrix elements of observables (namely, that their fluctuations cannot be completely random) becomes apparent when studying two-observable correlation functions~\cite {dalessio_quantum_2016}, and it has been a topic of much interest in the context of out-of-time-order correlators~\cite {foini_kurchan_19, chan_deluca_19, murthy_srednicki_19, brenes_pappalardi_21, wang_lamann_22}.}\BibitemShut {Stop}%
\bibitem [{\citenamefont {Zhang}\ \emph {et~al.}(2022)\citenamefont {Zhang}, \citenamefont {Vidmar},\ and\ \citenamefont {Rigol}}]{zhang_22}%
  \BibitemOpen
  \bibfield  {author} {\bibinfo {author} {\bibfnamefont {Y.}~\bibnamefont {Zhang}}, \bibinfo {author} {\bibfnamefont {L.}~\bibnamefont {Vidmar}},\ and\ \bibinfo {author} {\bibfnamefont {M.}~\bibnamefont {Rigol}},\ }\bibfield  {title} {\bibinfo {title} {Statistical properties of the off-diagonal matrix elements of observables in eigenstates of integrable systems},\ }\href {https://doi.org/10.1103/PhysRevE.106.014132} {\bibfield  {journal} {\bibinfo  {journal} {Phys. Rev. E}\ }\textbf {\bibinfo {volume} {106}},\ \bibinfo {pages} {014132} (\bibinfo {year} {2022})}\BibitemShut {NoStop}%
\bibitem [{\citenamefont {Pandey}\ \emph {et~al.}(2020)\citenamefont {Pandey}, \citenamefont {Claeys}, \citenamefont {Campbell}, \citenamefont {Polkovnikov},\ and\ \citenamefont {Sels}}]{pandey_claeys_20}%
  \BibitemOpen
  \bibfield  {author} {\bibinfo {author} {\bibfnamefont {M.}~\bibnamefont {Pandey}}, \bibinfo {author} {\bibfnamefont {P.~W.}\ \bibnamefont {Claeys}}, \bibinfo {author} {\bibfnamefont {D.~K.}\ \bibnamefont {Campbell}}, \bibinfo {author} {\bibfnamefont {A.}~\bibnamefont {Polkovnikov}},\ and\ \bibinfo {author} {\bibfnamefont {D.}~\bibnamefont {Sels}},\ }\bibfield  {title} {\bibinfo {title} {Adiabatic eigenstate deformations as a sensitive probe for quantum chaos},\ }\href {https://doi.org/10.1103/PhysRevX.10.041017} {\bibfield  {journal} {\bibinfo  {journal} {Phys. Rev. X}\ }\textbf {\bibinfo {volume} {10}},\ \bibinfo {pages} {041017} (\bibinfo {year} {2020})}\BibitemShut {NoStop}%
\bibitem [{\citenamefont {LeBlond}\ \emph {et~al.}(2021)\citenamefont {LeBlond}, \citenamefont {Sels}, \citenamefont {Polkovnikov},\ and\ \citenamefont {Rigol}}]{leblond_sels_21}%
  \BibitemOpen
  \bibfield  {author} {\bibinfo {author} {\bibfnamefont {T.}~\bibnamefont {LeBlond}}, \bibinfo {author} {\bibfnamefont {D.}~\bibnamefont {Sels}}, \bibinfo {author} {\bibfnamefont {A.}~\bibnamefont {Polkovnikov}},\ and\ \bibinfo {author} {\bibfnamefont {M.}~\bibnamefont {Rigol}},\ }\bibfield  {title} {\bibinfo {title} {Universality in the onset of quantum chaos in many-body systems},\ }\href {https://doi.org/10.1103/PhysRevB.104.L201117} {\bibfield  {journal} {\bibinfo  {journal} {Phys. Rev. B}\ }\textbf {\bibinfo {volume} {104}},\ \bibinfo {pages} {L201117} (\bibinfo {year} {2021})}\BibitemShut {NoStop}%
\bibitem [{\citenamefont {Kim}\ and\ \citenamefont {Polkovnikov}(2024)}]{kim_polkovnikov_24}%
  \BibitemOpen
  \bibfield  {author} {\bibinfo {author} {\bibfnamefont {H.}~\bibnamefont {Kim}}\ and\ \bibinfo {author} {\bibfnamefont {A.}~\bibnamefont {Polkovnikov}},\ }\bibfield  {title} {\bibinfo {title} {Integrability as an attractor of adiabatic flows},\ }\href {https://doi.org/10.1103/PhysRevB.109.195162} {\bibfield  {journal} {\bibinfo  {journal} {Phys. Rev. B}\ }\textbf {\bibinfo {volume} {109}},\ \bibinfo {pages} {195162} (\bibinfo {year} {2024})}\BibitemShut {NoStop}%
\bibitem [{\citenamefont {Foini}\ and\ \citenamefont {Kurchan}(2019)}]{foini_kurchan_19}%
  \BibitemOpen
  \bibfield  {author} {\bibinfo {author} {\bibfnamefont {L.}~\bibnamefont {Foini}}\ and\ \bibinfo {author} {\bibfnamefont {J.}~\bibnamefont {Kurchan}},\ }\bibfield  {title} {\bibinfo {title} {Eigenstate thermalization hypothesis and out of time order correlators},\ }\href {https://doi.org/10.1103/PhysRevE.99.042139} {\bibfield  {journal} {\bibinfo  {journal} {Phys. Rev. E}\ }\textbf {\bibinfo {volume} {99}},\ \bibinfo {pages} {042139} (\bibinfo {year} {2019})}\BibitemShut {NoStop}%
\bibitem [{\citenamefont {Chan}\ \emph {et~al.}(2019)\citenamefont {Chan}, \citenamefont {De~Luca},\ and\ \citenamefont {Chalker}}]{chan_deluca_19}%
  \BibitemOpen
  \bibfield  {author} {\bibinfo {author} {\bibfnamefont {A.}~\bibnamefont {Chan}}, \bibinfo {author} {\bibfnamefont {A.}~\bibnamefont {De~Luca}},\ and\ \bibinfo {author} {\bibfnamefont {J.~T.}\ \bibnamefont {Chalker}},\ }\bibfield  {title} {\bibinfo {title} {Eigenstate correlations, thermalization, and the butterfly effect},\ }\href {https://doi.org/10.1103/PhysRevLett.122.220601} {\bibfield  {journal} {\bibinfo  {journal} {Phys. Rev. Lett.}\ }\textbf {\bibinfo {volume} {122}},\ \bibinfo {pages} {220601} (\bibinfo {year} {2019})}\BibitemShut {NoStop}%
\bibitem [{\citenamefont {Murthy}\ and\ \citenamefont {Srednicki}(2019)}]{murthy_srednicki_19}%
  \BibitemOpen
  \bibfield  {author} {\bibinfo {author} {\bibfnamefont {C.}~\bibnamefont {Murthy}}\ and\ \bibinfo {author} {\bibfnamefont {M.}~\bibnamefont {Srednicki}},\ }\bibfield  {title} {\bibinfo {title} {Bounds on chaos from the eigenstate thermalization hypothesis},\ }\href {https://doi.org/10.1103/PhysRevLett.123.230606} {\bibfield  {journal} {\bibinfo  {journal} {Phys. Rev. Lett.}\ }\textbf {\bibinfo {volume} {123}},\ \bibinfo {pages} {230606} (\bibinfo {year} {2019})}\BibitemShut {NoStop}%
\bibitem [{\citenamefont {Brenes}\ \emph {et~al.}(2021)\citenamefont {Brenes}, \citenamefont {Pappalardi}, \citenamefont {Mitchison}, \citenamefont {Goold},\ and\ \citenamefont {Silva}}]{brenes_pappalardi_21}%
  \BibitemOpen
  \bibfield  {author} {\bibinfo {author} {\bibfnamefont {M.}~\bibnamefont {Brenes}}, \bibinfo {author} {\bibfnamefont {S.}~\bibnamefont {Pappalardi}}, \bibinfo {author} {\bibfnamefont {M.~T.}\ \bibnamefont {Mitchison}}, \bibinfo {author} {\bibfnamefont {J.}~\bibnamefont {Goold}},\ and\ \bibinfo {author} {\bibfnamefont {A.}~\bibnamefont {Silva}},\ }\bibfield  {title} {\bibinfo {title} {Out-of-time-order correlations and the fine structure of eigenstate thermalization},\ }\href {https://doi.org/10.1103/PhysRevE.104.034120} {\bibfield  {journal} {\bibinfo  {journal} {Phys. Rev. E}\ }\textbf {\bibinfo {volume} {104}},\ \bibinfo {pages} {034120} (\bibinfo {year} {2021})}\BibitemShut {NoStop}%
\bibitem [{\citenamefont {Wang}\ \emph {et~al.}(2022)\citenamefont {Wang}, \citenamefont {Lamann}, \citenamefont {Richter}, \citenamefont {Steinigeweg}, \citenamefont {Dymarsky},\ and\ \citenamefont {Gemmer}}]{wang_lamann_22}%
  \BibitemOpen
  \bibfield  {author} {\bibinfo {author} {\bibfnamefont {J.}~\bibnamefont {Wang}}, \bibinfo {author} {\bibfnamefont {M.~H.}\ \bibnamefont {Lamann}}, \bibinfo {author} {\bibfnamefont {J.}~\bibnamefont {Richter}}, \bibinfo {author} {\bibfnamefont {R.}~\bibnamefont {Steinigeweg}}, \bibinfo {author} {\bibfnamefont {A.}~\bibnamefont {Dymarsky}},\ and\ \bibinfo {author} {\bibfnamefont {J.}~\bibnamefont {Gemmer}},\ }\bibfield  {title} {\bibinfo {title} {Eigenstate thermalization hypothesis and its deviations from random-matrix theory beyond the thermalization time},\ }\href {https://doi.org/10.1103/PhysRevLett.128.180601} {\bibfield  {journal} {\bibinfo  {journal} {Phys. Rev. Lett.}\ }\textbf {\bibinfo {volume} {128}},\ \bibinfo {pages} {180601} (\bibinfo {year} {2022})}\BibitemShut {NoStop}%
\end{thebibliography}%
\end{document}